\documentclass[aps,twocolumn,prl,superscriptaddress,amsmath,amssymb,amsfonts]{revtex4-1} 

\usepackage[T1]{fontenc}
\usepackage[latin9]{inputenc}
\usepackage{lmodern} % load a font with all the characters
\usepackage{color}
\usepackage{bbold}
\usepackage{dsfont}
\usepackage{graphicx}
\usepackage[caption=false]{subfig}
\usepackage[colorlinks]{hyperref}
\begin{document}

\global\long\def\eqn#1{\begin{align}#1\end{align}}
\global\long\def\ket#1{\left|#1\right\rangle }
\global\long\def\bra#1{\left\langle #1\right|}
\global\long\def\bkt#1{\left(#1\right)}
\global\long\def\sbkt#1{\left[#1\right]}
\global\long\def\cbkt#1{\left\{#1\right\}}
\global\long\def\abs#1{\left\vert#1\right\vert}
\global\long\def\der#1#2{\frac{{d}#1}{{d}#2}}
\global\long\def\pard#1#2{\frac{{\partial}#1}{{\partial}#2}}
\global\long\def\re{\mathrm{Re}}
\global\long\def\im{\mathrm{Im}}
\global\long\def\dd{\mathrm{d}}
\global\long\def\ddd{\mathcal{D}}

\global\long\def\avg#1{\left\langle #1 \right\rangle}
\global\long\def\mr#1{\mathrm{#1}}
\global\long\def\mb#1{{\mathbf #1}}
\global\long\def\mc#1{\mathcal{#1}}
\global\long\def\tr{\mathrm{Tr}}
\global\long\def\dbar#1{\Bar{\Bar{#1}}}

\global\long\def\nth{$n^{\mathrm{th}}$\,}
\global\long\def\mth{$m^{\mathrm{th}}$\,}
\global\long\def\non{\nonumber}

\newcommand{\orange}[1]{{\color{orange} {#1}}}
\newcommand{\cyan}[1]{{\color{cyan} {#1}}}
\newcommand{\blue}[1]{{\color{blue} {#1}}}
\newcommand{\yellow}[1]{{\color{yellow} {#1}}}
\newcommand{\green}[1]{{\color{green} {#1}}}
\newcommand{\red}[1]{{\color{red} {#1}}}
\global\long\def\todo#1{\yellow{{$\bigstar$ \orange{\bf\sc #1}}$\bigstar$} }
\title{Non-Markovian collective emission from macroscopically separated emitters}

\begin{abstract}
We study the collective radiative decay of a system of two two-level emitters coupled to a one-dimensional waveguide in a regime where their separation is comparable to the coherence length of a spontaneously emitted photon. The electromagnetic field propagating in the cavity-like geometry formed by the emitters exerts a retarded backaction on the system leading to strongly non-Markovian dynamics. The collective spontaneous emission rate of the emitters exhibits an enhancement or inhibition beyond the usual Dicke super- and sub-radiance due to a self-consistent coherent time-delayed feedback.
\end{abstract}

\author{Kanupriya Sinha}
\email{kanu@umd.edu}
\affiliation{US Army Research Laboratory, Adelphi, Maryland 20783, USA}
\affiliation{Joint Quantum Institute, University of Maryland, College Park, MD 20742, USA}
\author{Pierre Meystre}
\affiliation{Department of Physics and College of Optical Sciences, University of Arizona, Tucson, AZ 85721, USA}
\author{Elizabeth A. Goldschmidt}
\affiliation{US Army Research Laboratory, Adelphi, Maryland 20783, USA}
%\affiliation{Joint Quantum Institute, University of Maryland, College Park, MD 20742, USA}
\author{Fredrik K. Fatemi}
\affiliation{US Army Research Laboratory, Adelphi, Maryland 20783, USA}
\author{S. L. Rolston}
\affiliation{Joint Quantum Institute, University of Maryland, College Park, MD 20742, USA}
\affiliation{Department of Physics, University of Maryland, College Park, MD 20742, USA}
\author{Pablo Solano}
\email{solano.pablo.a@gmail.com}
\affiliation{Department of Physics, MIT-Harvard Center for Ultracold Atoms, and Research Laboratory of Electronics, Massachusetts Institute of Technology, Cambridge, MA 02139, USA}
\maketitle

{\it Introduction.---}{Long-distance interactions are a central tenet of many quantum systems and processes, including large-scale quantum networks, distributed quantum sensing and information processing~\cite{Kimble08, Schoelkopf08, Pichler17, Pichler16, Ge2018}}. When the separations between emitters become comparable to the coherence length of the photons mediating their interaction, interference properties of the electromagnetic (EM) field can be modified due to retardation. In such cases, the backaction of the EM field on the emitters leads to a coherent time-delayed feedback on the system dynamics \cite{Whalen17, Grimsmo15}, thus rendering it non-Markovian \cite{BPbook, Breuer16, deVega17, Fleming12}.

Non-Markovian open system dynamics can have a variety of physical origins such as structured bath spectral densities, strong system-bath couplings, low temperatures, or initial system-bath correlations among others \cite{Breuer16, deVega17,HPZ92, HPZ93, Vasile14, Groblacher15}. The effects of non-Markovianity have  been investigated in collective atomic states in the context of structured reservoirs \cite{Thanopulos19, Vats98, John97, AGT2017PRA, AGT2017PRL} and in the strong-coupling regime \cite{Daniele19}. Furthermore, delay-induced non-Markovian dynamics has been previously shown in the context of spontaneous emission of single atoms \cite{Giessen96,Dorner02,Tufarelli13, Tufarelli14, Cook87,Carmele13, Beige02, Guimond16, Guimond17}, bound states in continuum (BIC) of the EM field \cite{Hsu16, Fong17, Calajo19, Facchi19, Dinc18}, and entanglement generation in emitters coupled to waveguides \cite{ Pichler16, Carlos13}. 

{Additional insights into non-Markovian effects in such a regime can be gained from studying the simple, yet rich quantum optics phenomenon of collective spontaneous emission of two two-level emitters. Cooperative effects in spontaneous emission  have an extensive historical background \cite{Dicke, Haroche, Rehler71, Eberly72}, and have been  experimentally observed across a range of physical systems~\cite{Skribanowitz, Gross76, Pavolini85, Devoe96, Scheibner07, Mlynek14, Solano2017, Kim2018, Chen2018}. While the influence of retardation on these effects has been previously studied in Refs. \cite{Milonni74, Arecchi70}, the non-Markovian dynamics emerging in macroscopically delocalized  collective systems is yet unexplored.}

In this letter we study the collective radiative dynamics of a pair of macroscopically separated emitters and show that it exhibits non-Markovian features caused by a self-consistent coherent  time-delayed feedback.  We specifically  consider here the emitters are prepared in a super- or sub-radiant  electronic state, and present an exact analytical solution of the dynamics of the collective spontaneous emission. We demonstrate that the retarded backaction of the EM field on the emitters can lead to a further enhanced (inhibited) spontaneous emission rate for superradiant (subradiant) states beyond the usual Dicke superradiance (subradiance) \cite{Dicke, Haroche}. 
\begin{figure}[b]
    \centering
    \includegraphics[width = 2.75 in]{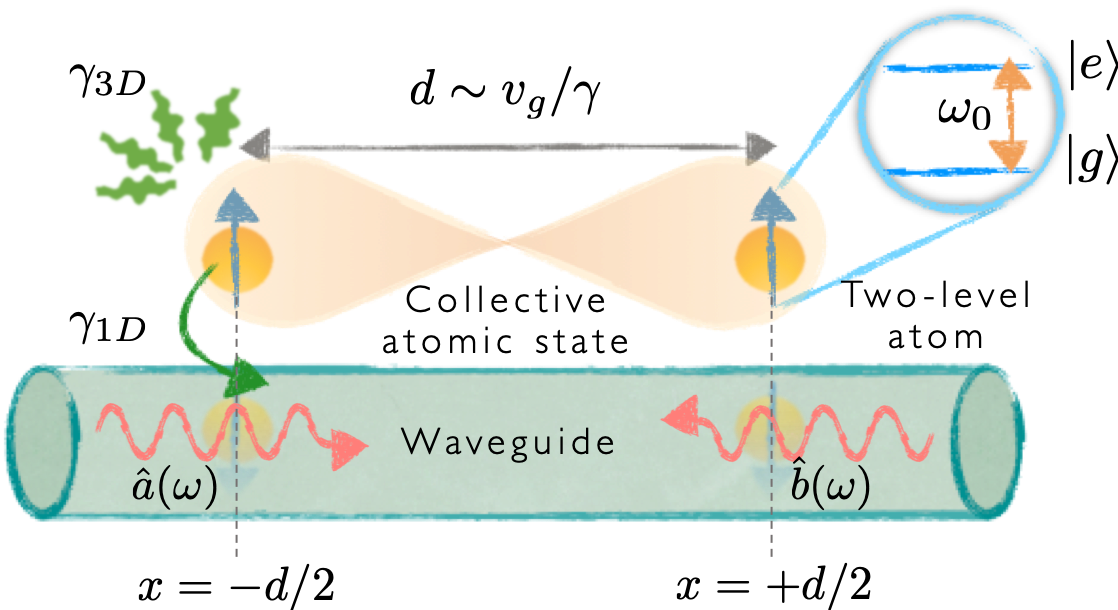}
    \caption{Two two-level emitters prepared in a collective state coupled to an optical waveguide. The emitters are located at positions $x_{1,2} = \pm d/2$, with $d$  comparable to the coherence length $\sim v_g/\gamma$. The rates $\gamma_{3D}$ and $\gamma_{1D}$ refer to the emitter spontaneous emission rates  into free space and guided modes respectively. The mode operators $\hat{a}(\omega)$ and $\hat{b}(\omega)$ refer to annihilation operators for the  right- and left-propagating waveguide modes respectively.}
    \label{Schematic}
\end{figure}

 We consider two two-level emitters coupled to a waveguide are separated by a distance $d$ comparable to the coherence length ${\sim v_g/\gamma}$ of a spontaneously emitted photon, with $v_{g}$ being the group velocity of the field and $\gamma$ the spontaneous emission rate of the individual emitters (see Fig.\,\ref{Schematic}). To gain an intuitive understanding of the non-Markovian nature of this system, consider the following apparent ``superradiance paradox:'' Assume that the distance $d$ between two emitters prepared in a superradiant state is smaller than the coherence length of an independently emitted photon, but larger than that of a superradiant photon, $  v_{g}/\gamma>d>v_{g}/(2\gamma)$. Given that superradiance is an interference effect, one would expect to observe superradiant emission if there is no way to distinguish which atom emitted the field \cite{FicekTanas}. Now if the emitters radiate collectively, with an emission rate $2\gamma$, then the coherence length of the emitted photons ($v_{g}/(2\gamma)$) is too short to allow for the fields radiated by the two emitters to interfere, suggesting that they should have emitted independently. On the other hand, if we assume that they do emit independently, then the coherence length of the emitted photons ($v_{g}/\gamma$) is long enough  that there should be interference and as a result the emitters should emit at the superradiant rate of $2\gamma$ instead. This seeming paradox points to the failure of the Markov approximation: the conventional notion of an exponential decay defining the photon coherence length is no longer valid, and it is necessary instead to consider a full non-Markovian treatment of the system dynamics.

{\it Formal development.---} {The total Hamiltonian for the emitters+ field system is $H = H_E + H_F + H_\mr{int}$, where $ H_E  = \hbar \omega_0\sum_{m = 1,2} \hat {\sigma}_+^{(m)}\hat {\sigma}_-^{(m)} $ is the free Hamiltonian for the emitters of resonance frequency $\omega_0$, with  $\hat {\sigma}_+^{(m)}$ is the creation operator of an excitation in the $m^{th}$ emitter; $H_F = \hbar \omega \int _0 ^\infty \dd\omega\sbkt{ \hat{a}^\dagger\bkt{\omega}\hat{a}\bkt{\omega} + \hat{b}^\dagger\bkt{\omega}\hat{b}\bkt{\omega}}$ is the free Hamiltonian for the field, with $\hat {a}\bkt{\omega}$ and $\hat {b}\bkt{\omega}$ the annihilation operators for the right- and left-propagating field modes of the waveguide respectively; and $H_\mr{int }$ is the emitter-field interaction Hamiltonian.}

We proceed by making the electric-dipole and rotating-wave approximations (RWA) and expressing the emitters-field interaction Hamiltonian in the interaction picture  with respect to the total free Hamiltonian $H_E + H_F$ as
\begin{eqnarray}
    \label{hint} 
    H_{\mr {int}} = &\hbar \sum _{m = 1}^2 \int_0 ^\infty \dd\omega \left[ g(\omega) \hat {\sigma}_+^{(m)}\left\{\hat{a}(\omega) e^{i \omega x_m/v_g} \right.\right. \nonumber\\
     &\left.\left.+\hat{b}(\omega) e^{-i \omega x_m /v_g}\right\}e^{-i (\omega - \omega_0 )t}+ {\rm h.c.} \right] ,
\end{eqnarray}
where $g(\omega)$ is the atom-field coupling strength \cite{footnoteop, footnoteg}. To isolate the non-Markovian behavior arising from the retardation effects from that due to a  structured reservoir we assume a flat spectral density of the field modes around the  resonance of the emitters such that $g\bkt{\omega}\approx g\bkt{\omega_0} $. 

Assuming that the total emitters plus field system is initially prepared in the single-excitation manifold, and considering that in the RWA the Hamiltonian preserves the total number of excitations, the state at time $t>0$ is
\eqn{\label{psit}
&\ket{\Psi \bkt{t}} = \sum _{m = 1}^2 c_m\bkt{t} \hat{\sigma}_+ ^{(m)} \ket{g,g,\cbkt{0}}+ \non\\
&\int_0 ^\infty \dd\omega \sbkt{c_a\bkt{ \omega, t} \hat{a}^\dagger \bkt{\omega}+ c_b\bkt{ \omega, t} \hat{b}^\dagger \bkt{\omega}}\ket{g,g,\cbkt{0}},
}
where $c_m $ and $c_{a,b } (\omega)$ are the excitation amplitudes for the $m^{\mr{th}}$ emitter and the guided field modes with frequency $\omega$ respectively, and  $\ket{g,g,\cbkt{0}}$ is the ground state of the total system, with $\ket{\cbkt{0}}$ the field vacuum state.  Tracing out the field modes, evolution of emitter excitation amplitudes is given by \cite{Asl03}
\eqn{\label{eqmot}
\dot{c}_m\bkt{t} = -\frac{\gamma}{2} \sbkt{c_m \bkt{t}  + \beta c_n \bkt{t - d/v_g} \Theta \bkt{ t - d/v_g} e^{i \phi_p} }.
}
% \eqn{
% \dot{c}_1\bkt{t} = -\frac{\gamma}{2} \sbkt{c_1 \bkt{t}  + \beta c_2 \bkt{t - d/v_g} \Theta \bkt{ t - d/v_g} e^{i \phi_0} },\non\\
% \dot{c}_2\bkt{t} = -\frac{\gamma }{2}\sbkt{c_2 \bkt{t}  + \beta c_1 \bkt{t - d/v_g} \Theta \bkt{ t - d/v_g} e^{i \phi_0} },
% }
for $m\neq n$, where $\phi_p \equiv k_0 d = 2 p \pi$ is the field phase difference upon propagation, which we assume to be an integer multiple of $2 \pi$, $\gamma \equiv \gamma_{1D} + \gamma_{3D}$ is the total spontaneous emission rate, and $\gamma_{1D} = \beta \gamma \equiv 4\pi \abs{g\bkt{\omega_0}}^2$ is the spontaneous emission rate into the waveguide, with $\beta$ the coupling efficiency of the emitters to the waveguide. The second term in Eq.\,\eqref{eqmot} represents the retarded backaction of the other emitter via the field with a delay $t = d/v_g$. 

For emitters initially in the super- or sub-radiant states $\ket{\Psi_{\substack{\mr{sup}\\ \mr{sub}}}} \equiv \frac{1}{\sqrt{2}}\bkt{\ket{eg}\pm\ket{ge}}$ \cite{footnote1}, one can write the Laplace transformed coefficients $\tilde c_m\bkt{s} \equiv \int_0 ^\infty \dd t\, e^{- s t} c_m \bkt{t}$ as
\eqn{\label{cssup}
\tilde {c}_{\mr{sup}}\bkt{s}
= & \frac{1}{\sqrt{2}\gamma\sbkt{\tilde{s}+1/2+\beta e^{ -\eta \tilde{s}}/2 }},\\
\label{cssub}
\tilde {c}_{\mr{sub}}\bkt{s} =& \frac{1}{\sqrt{2}\gamma\sbkt{\tilde{s}+1/2-\beta e^{ -\eta \tilde{s}}/2 }},
% \eqn{\label{cssup}
% \tilde {c}_{1}^{\mr{sup}}\bkt{s}  = {c}_{2}^{\mr{sup}}\bkt{s} 
% = & \frac{1}{\sqrt{2}\gamma\sbkt{\tilde{s}+1/2+\beta e^{ -\eta \tilde{s}}/2 }},\\
% \label{cssub}
% \tilde {c}^{\mr{sub}}_{1}\bkt{s} =- {c}^{\mr{sub}}_{2}\bkt{s} 
% = & \frac{1}{\sqrt{2}\gamma\sbkt{\tilde{s}+1/2-\beta e^{ -\eta \tilde{s}}/2 }},
}
where  $\tilde s \equiv s/\gamma $ and $\eta\equiv d \gamma/v_g$ is the separation between the emitters normalized by the photon coherence length. Here $\tilde {c}_{\mr{sup}}=\tilde {c}^{\mr{sup}}_1=\tilde {c}^{\mr{sup}}_2$ and $\tilde {c}_{\mr{sub}}=\tilde {c}^{\mr{sub}}_1=-\tilde {c}^{\mr{sub}}_2$ are the Laplace space  probability amplitudes for the super- and sub-radiant cases respectively \cite{footnote2}.

Consider next the case where the emitters are slightly separated,  $\eta\ll 1$. Up to linear terms in $\eta $ 
\eqn{
\tilde {c}_{\substack{\mr{sup}\\\mr{sub}}}\bkt{\tilde s} &\approx \frac{1}{\sqrt{2} \sbkt{s(1 \mp \beta \eta/2) +\gamma/2 \bkt{1\pm\beta  } }},}  
which yields an effective spontaneous emission rate
\eqn{\gamma^{\substack{\mr{sup}\\\mr{sub}}}\approx\frac{{1\pm \beta  }}{1 \mp\beta \eta/2}\gamma.}
For a small but finite delay $0<\eta\ll1$, this can potentially exceed the usual Dicke superradiant emission rate of $2\gamma$ for $\beta = 1$. Also, for a subradiant state with an imperfect coupling  ($\beta<1$), the effective decay  for slightly separated emitters can be slower than that for coincident ones. This  surprising enhancement and inhibition of the collective spontaneous emission can be attributed to  a stimulated emission as  the correlated field emitted from one of the emitters interferes  with that from the other \cite{Cray82}. The separation dependence of the  collective emission rate in addition to the phase difference, demonstrates the influence of retardation on the interference.

We now consider the general case of arbitrarily separated emitters, for which we present an exact analytical solution of the equations of motion~(\ref{eqmot})  based on a well-developed mathematical treatment of delay differential equations (see \cite{Corless96} and the Supplemental Material (SM) \cite{SM} for details). The general expression for the excitation amplitudes of the emitters is
\eqn{\label{eq:Cs}
c_{\substack{\mr{sup}\\\mr{sub}}}(t) = \frac{1}{\sqrt{2}}\sum_{n \in \mathbb{Z}}  \alpha_n^{(\pm)} e^{ -\gamma_n^{(\pm)}  t/2},}
where $\alpha_n^{(\pm)} \equiv \sbkt{1+ W_{n} \bkt{\mp\frac{\eta}{2}\beta  e^{\eta/2 } } }^{-1}$ and the effective decay rate $\gamma_ n ^{(\pm)}\equiv \gamma \sbkt{1-   2 W_{n} \bkt{\mp\frac{\eta}{2}\beta  e^{ \eta/2 } }/\eta }$, with $W_n (x)$ the $n^\mr{th}$ branch of the Lambert $W$-function, which is commonly used to describe systems that exhibit time-delayed feedback \cite{Corless96, Asl03}. We now discuss the consequences of this analytical solution, which is the main result of this work.

\begin{figure}[t]
    \centering
    \subfloat[]{\includegraphics[width = 3 in]{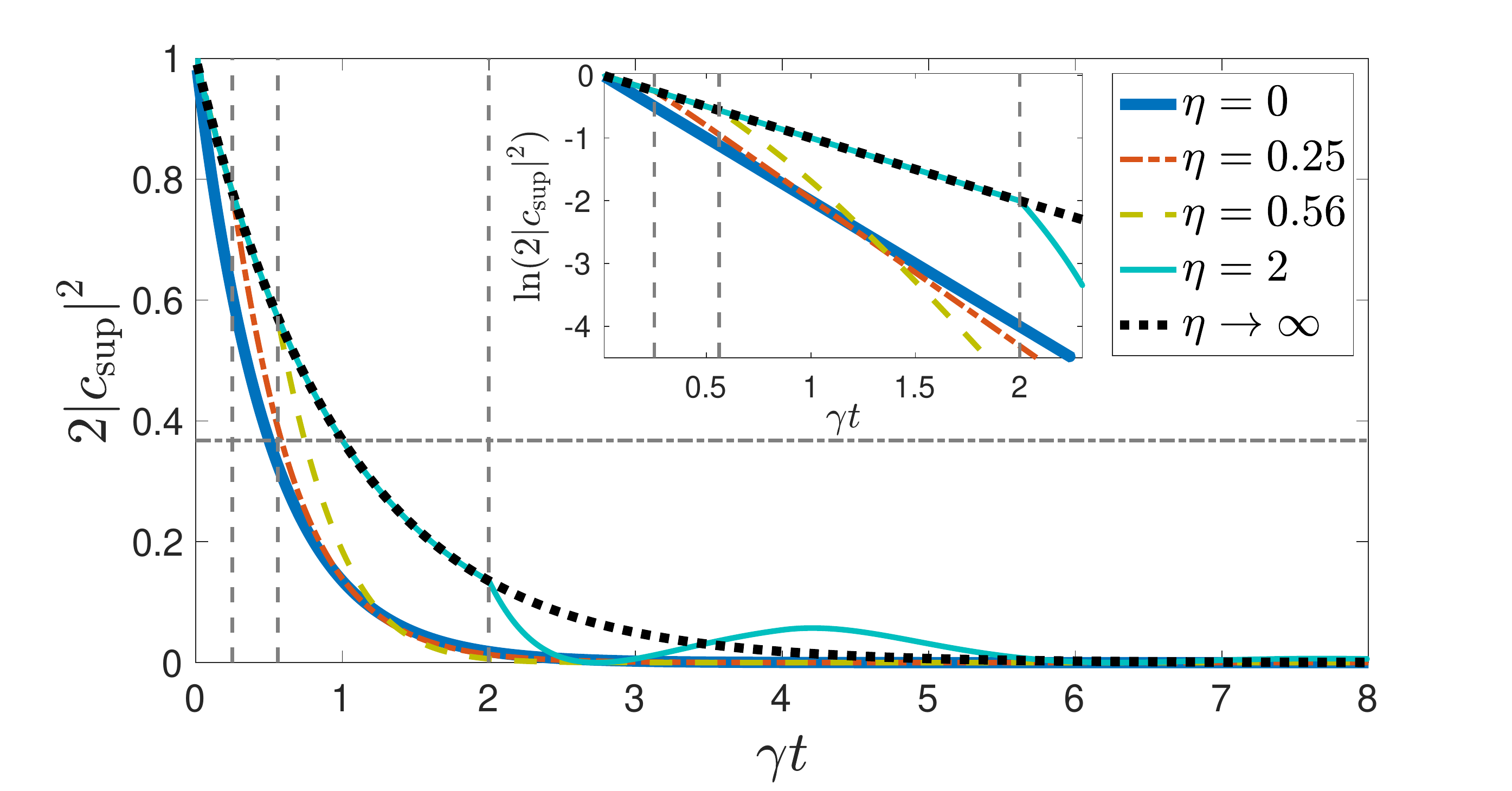}}\\
       \subfloat[]{\includegraphics[width = 3 in]{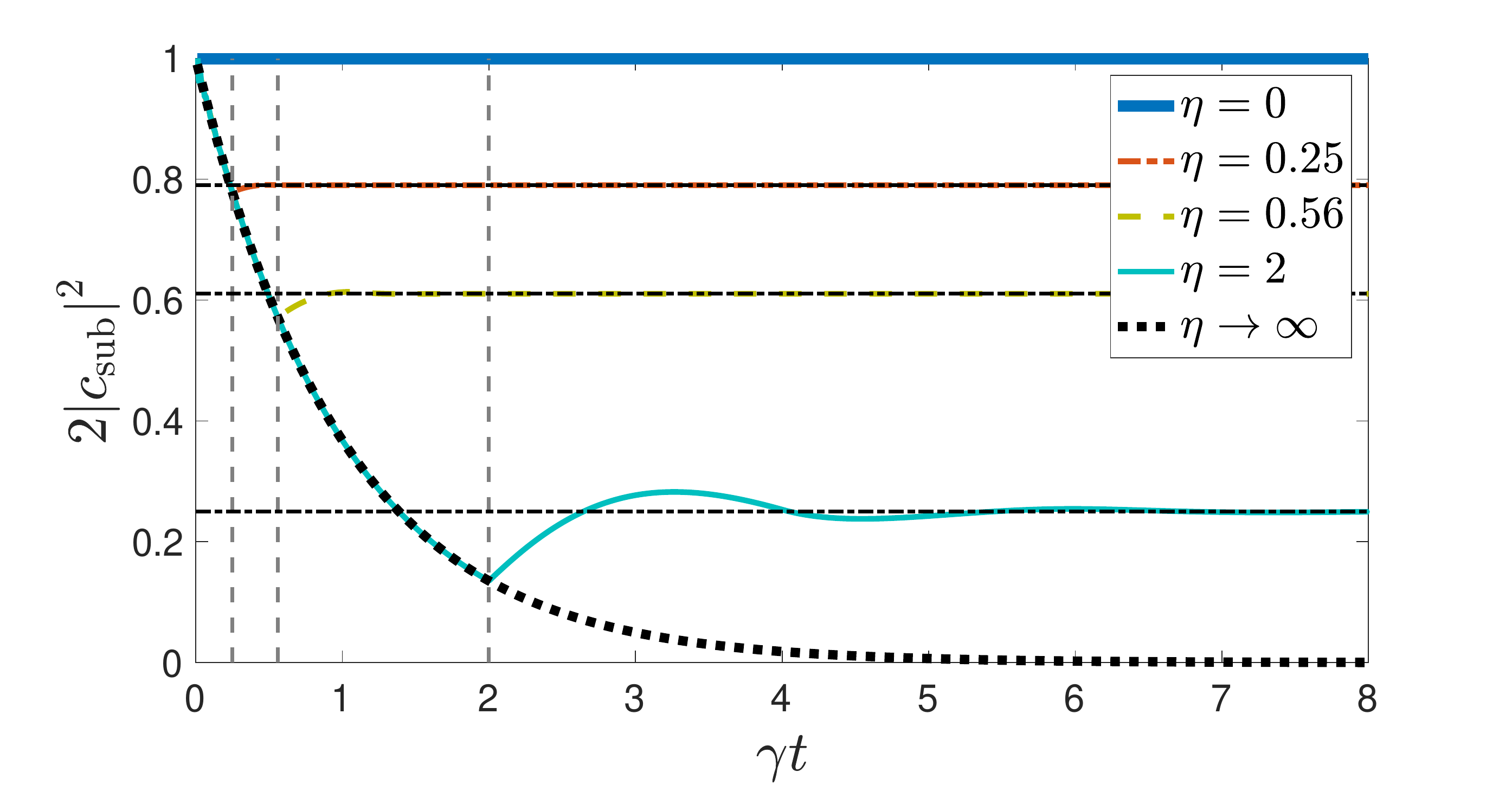}}
    \caption{Emitters excitation probabilities as a function of time and for different separations $\eta$ for initial (a) superradiant and (b) subradiant states, assuming perfect coupling efficiency ($\beta = 1$). The solid and dotted curves depict the dynamics for  ${\eta =0}$ and  ${\eta =\infty}$ respectively. For intermediate emitter separations the emitters decay independently with a rate $\gamma$ until $\gamma t  = \eta$ (indicated by the dashed vertical lines), and collectively afterwards. For the critical separation $\eta\approx\eta_c\approx 0.56 $, we observe an instantaneous superradiant spontaneous emission rate of $\gamma_\mr{inst}\approx 4.59\gamma$.  The $1/e$ value of the initial emitters excitation probability is reached first for coincident  emitters (depicted by the gray horizontal dashed-dotted line). In the subradiant case the emitter excitation probability reaches the asymptotic value $(1+\eta/2)^{-2}$, shown by the horizontal dashed-dotted lines. }
    \label{Fig2}
\end{figure}

\begin{figure}
    \centering
    \includegraphics[width = 3 in]{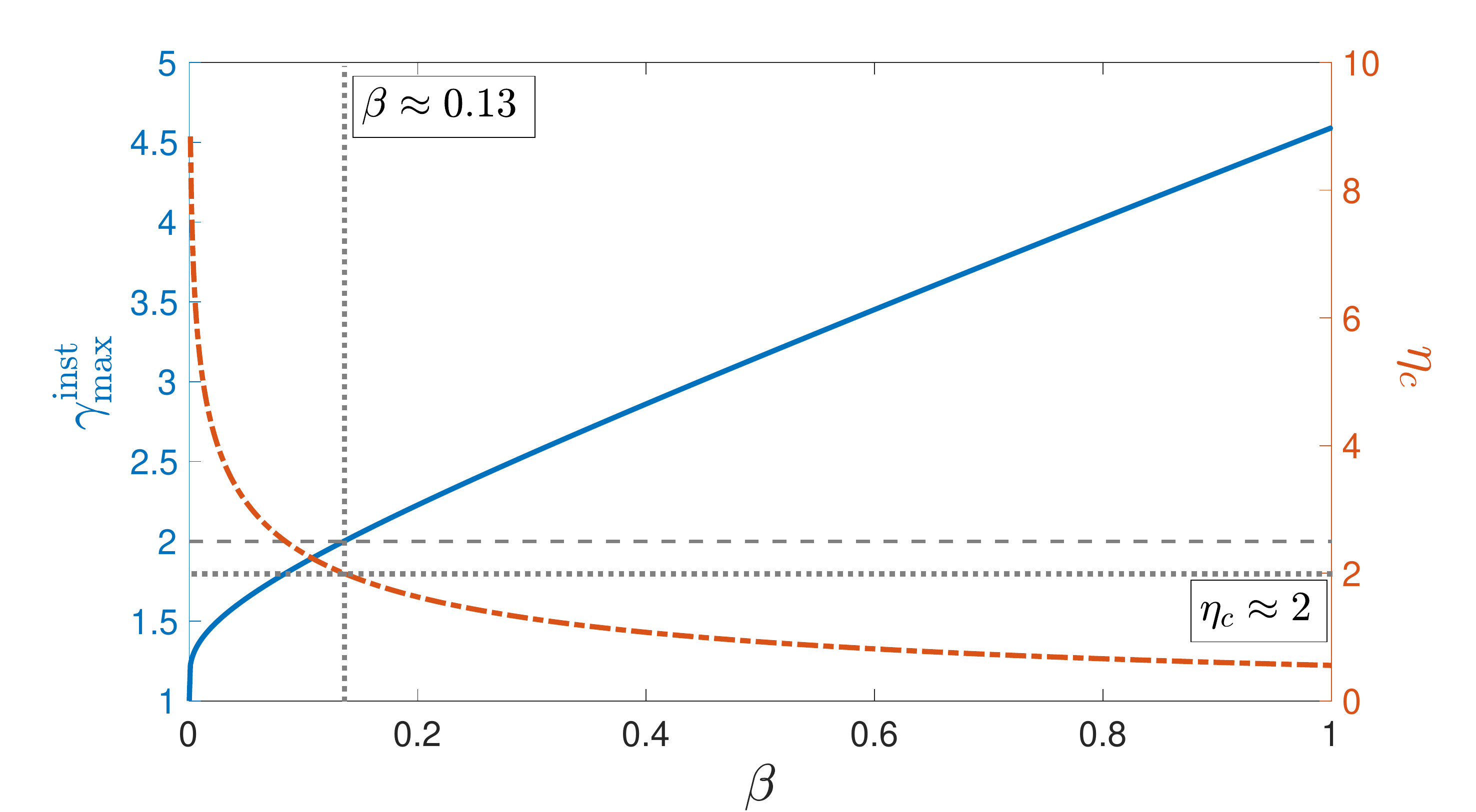}
    \caption{Instantaneous decay rate $\gamma_{\mr{inst }}^\mr{max}$ (blue solid line) and associated critical emitter separation $\eta_c$ (red dashed-dotted line) as a function of the emitter-waveguide coupling efficiency $\beta$ for an initial superradiant state. The horizontal dashed line depicts an instantaneous collective emission rate of $2\gamma$, which corresponds to a coupling efficiency of $\beta \approx 0.13$. This illustrates that the collective emission rate of $2\gamma$ of usual Dicke superradiance can be exceeded for sufficiently large emitter-waveguide coupling efficiency and appropriate emitter separations.}
    \label{Fig3a}
\end{figure}

{\it Results.---}Consider first the dynamics of a superradiant initial state. From Eq.\,\eqref{eq:Cs} and the properties of the Lambert-$W$ function one finds that the superradiant solution has imaginary exponents for $\eta>\eta_c$, where we have introduced the normalized critical distance $\eta_c \equiv 2W_0\bkt{1/(e\beta)}$ \cite{Asl03}. Thus for $\eta\geq\eta_c$, the atomic excitation amplitudes exhibit oscillations as the atoms decay to their ground state. These can be understood in terms of a field wavepacket bouncing back and forth between the emitters \cite{Milonni74, SM}. For $\beta=1$ this occurs for separations $d> 0.56 v_g/\gamma$, as shown in Fig\,\ref{Fig2}. 

For separations $\eta<\eta_c$ the emitters radiate independently until a time $\gamma t =\eta $ and collectively afterwards, with an instantaneous decay rate given by
\eqn{ \label{gammainst}\gamma_{\mr{inst }} =\gamma\sbkt{ 1 - \frac{W_0\bkt{-\frac{\eta}{2} \beta e^{ \eta /2}} }{\eta/2}}.}
For a given value of $\beta$, this rate reaches a maximum $\gamma_{\mr{inst }}^\mr{max}$ when the normalized emitter separation equals its critical value $\eta = \eta_c $, with ${\gamma_{\mr{inst }}^\mr{max}/\gamma  = 1 - \frac{W_0\bkt{-1/e} }{W_0\bkt{1/(e\beta)}}}$, as shown in Fig.\,\ref{Fig3a}. In the absence of losses and for perfect emitter-waveguide coupling efficiency $(\beta = 1)$ the maximum instantaneous spontaneous emission rate is $ \gamma_{\mr{inst }}^\mr{max}/\gamma  \approx 4.59$, in stark contrast with superradiant emission in Markovian systems.
%  Note that despite the fact that the emitters can radiate at some times at more than twice the superradiant rate and can reach the ground state in a shorter time, it is in the case of usual Dicke superradiance $\bkt{\eta =0}$ that they decay the fastest in terms of reaching the $1/e$ fraction of the initial population, see Fig.\,\ref{Fig2}(a)

In the case of a subradiant initial state, for $\beta = 1$, the steady state of the dynamics corresponds to a BIC \cite{Calajo19}. The probability of reaching the BIC starting initially in the subradiant state of the atoms is given by ${\abs{ \avg{\Psi\bkt{t\rightarrow\infty} \vert\Psi_{\mr{BIC}}}}^2 = 1/\bkt{1+\eta/2}}$ \cite{SPIE}. The total probability of  the emitters being excited in the steady state as $ \abs {c_{1,2}^\mr{sub}(\infty)}^2 \rightarrow \frac{1}{2\bkt{{1+ \eta/2}}^2}$, see Fig.\,\ref{Fig2}\,(b). {We also note that for an initial subradiant state with a delay of $\eta \approx 0.8$, it is possible to achieve a maximal emitter-field steady state entanglement \cite{SM}.}

\begin{figure}[ht]
    
\centering\includegraphics[width = 3.3 in]{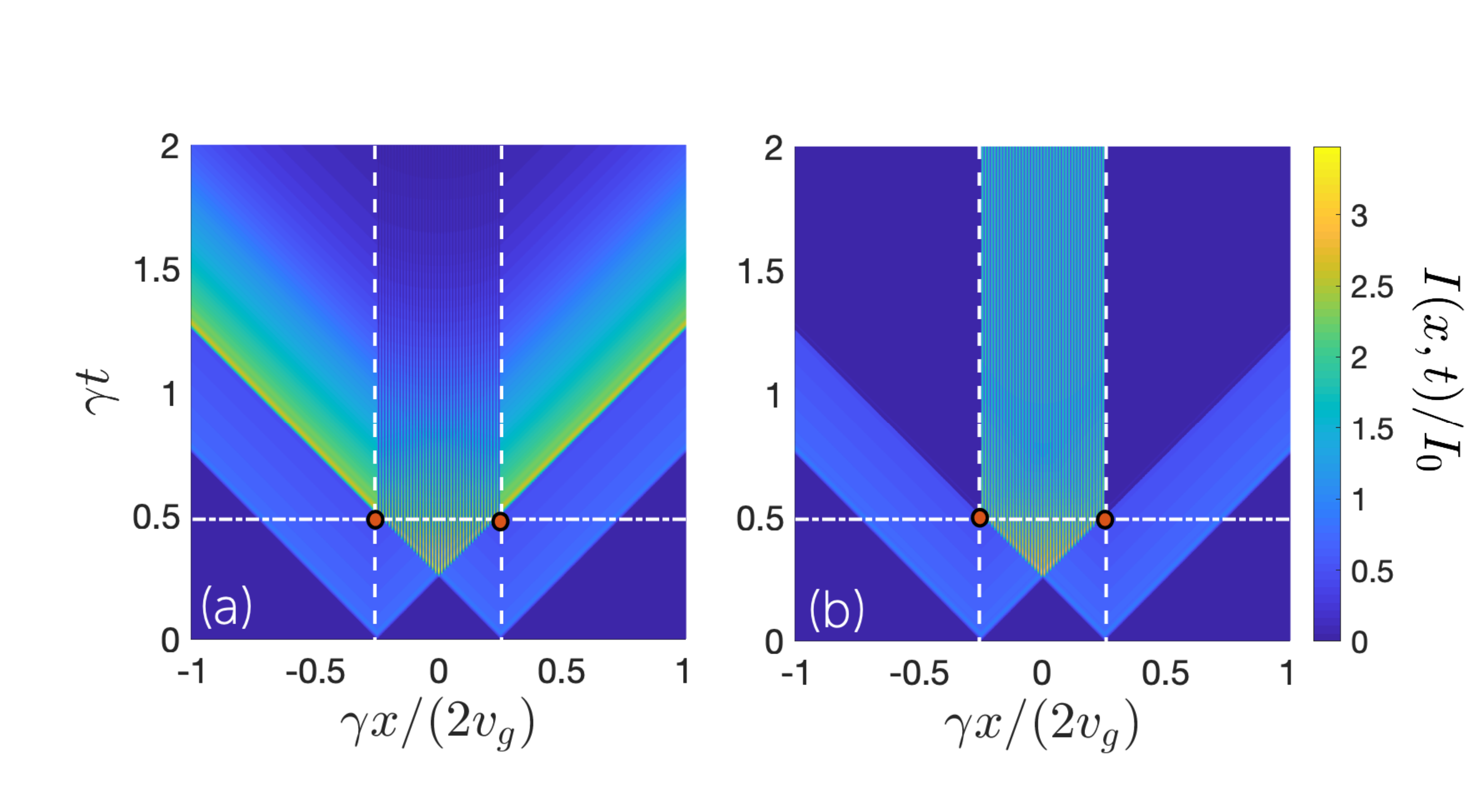}
\caption{Normalized field intensity as a function of position and time for $\eta = 0.5$ and (a) superradiant and (b) subradiant initial state. The positions of emitters at $\gamma x_{1,2}/(2v_g)\approx \pm 0.25$ are depicted by the vertical dashed line. As the field from one of the emitters reaches the other at $\gamma t = \eta$ (dashed-dotted line), their interference results in either a collective radiation burst or reflection of the field into the cavity formed by the emitters. }
    \label{Fig4}
\end{figure}

It is also instructive to explore the cooperative nature of the atom-field dynamics from the perspective of the emitted field intensity $ I\bkt{x,t} \propto \avg{\Psi\bkt{t}\vert\hat{E}^\dagger\bkt{x,t} \hat E\bkt{x,t}\vert\Psi\bkt{t}} $, where the electric field operator is $\hat{E}\bkt{x,t} \propto \int_0^\infty  \dd k \sbkt{\hat{a} \bkt{k}e^{i k x} + \hat{b} \bkt{k}e^{-i k x}}e^{i \omega t}$ and $|\Psi(t)\rangle$ the state of the system (see \cite{SM} for details). Fig.\,\ref{Fig4} shows that the fields emitted by the two emitters in the superradiant (subradiant) case interfere constructively (destructively) when the light cones of the two emitters reach each other. Thereafter, depending on their relative phase they produce an interference pattern that is either constructive, leading to a collective `{superduperradiant}' burst with an instantaneous emission rate greater than $2\gamma$, or destructive, resulting in the a perfect reflection of the field into the optical cavity created by the two atoms. {The non-exponential decay of the emitters is an unambiguous signature of the non-Markovian evolution of the system, a result of the self-consistent backaction of the EM field bath, which is accounted for, by a departure from the usual Lindblad dynamics. We further quantify the non-Markovianity of the system in the Supplemental Material [68], which shows that the system is non-Markovian for any value of $\eta$,  approaching a Markovian behavior for $\eta\rightarrow 0$.} 

{Noting that in the presence of delay the instantaneous collective decay rate can exceed  that of standard Dicke superradiance, one might wonder if the total collective emission into the waveguide also gets enhanced. An important figure of merit to quantify the collective nature of the system in this regard is its cooperativity } $\mc{C} \equiv\gamma_\mr{in}/\gamma_{3 D}$ \cite{PabloReview},
such that $\gamma_\mr{in} = \lim_{t\rightarrow\infty} \int_0^\infty \dd\omega\,\sbkt{ \abs{c_a\bkt{\omega, t}}^2 + \abs{c_b\bkt{\omega, t}}^2
}$ is the fraction of the field emitted into the waveguide and $\gamma_{3D} = \gamma(1 - \beta)$ is the fraction of the field that escapes out to the non-guided modes \cite{footnote4}. This can be evaluated as \cite{SM}
\eqn{
\mc{C}_{\substack{\mr{sup}\\\mr{sub}}}=& \frac{\beta}{1 - \beta} \sum_{m,n} \frac{\alpha_n^{(\pm)}{\alpha_m^{(\pm)}}^\ast}{ \gamma_n^{(\pm)}+  {\gamma_m^{(\pm)}}^\ast} \non\\
&\sbkt{2 \pm \cbkt{e^{- \eta\gamma_n^{(\pm)}  /(2\gamma)} + e^{- \eta{\gamma_m^{(\pm)}}^\ast /(2\gamma)}}.}
% \gamma_{\mr{in}}^{\mr{sub}} = \gamma\beta \sum_{m,n} \frac{\alpha_n^{(-)}{\alpha_m^{(-)}}^\ast}{ \gamma_n^{(-)}+  {\gamma_m^{(-)}}^\ast} \sbkt{2 - \cbkt{e^{- \eta\gamma_n^{(-)}  /(2\gamma)} + e^{- \eta{\gamma_m^{(-)}}^\ast /(2\gamma)}}}
}
For $\eta >0 $, the cooperativity for a superradiant state is reduced compared to that of coincident emitters $(\eta = 0)$ as the total collective emission into the guided modes decreases with the emitter separation. In contrast, for an anti-symmetric state we find an enhanced emission into the waveguide as $\eta $ is increased. This is due to the emission of the field into guided modes by the individual emitters until $\gamma t = \eta$, before they start acting collectively (see Fig.\,\ref{Fig4}\,(b)). {Given that cooperativity is an important figure of merit in quantum information applications, this result illustrates that retardation effects need to be carefully considered in quantum network protocols based on long distance emitters~\cite{Kurizki15}.}

\textit{Summary and outlook.---} We have shown that the collective radiative decay of two emitters coupled to a one-dimensional waveguide  is subject to non-Markovian modifications due to the time-delayed backaction of the electromagnetic field upon the emitters. When prepared in a superradiant initial state they can exhibit time-dependent decay rates that can instantaneously surpass the standard Dicke superradiance rate. The system also allows for  long-lived subradiant states characterized by a bound state in the field trapped in the region between the emitters. These effects can be understood as a combination of Dicke super- or subradiance and a retardation of the field wavepacket where the electromagnetic field senses its boundary conditions with a significant delay. 

A key parameter for characterizing the dynamics is the emitter separation relative to the photon coherence length $\eta\equiv d \gamma/v_g$. It captures the combined physical origin of non-Markovian behavior, as an appreciable value of $\eta $ can be achieved by increasing the emitter separation $d$, but also by increasing the system-environment coupling as in \cite{Carlos13} or by exploiting slow group velocities achievable in the presence of a band gap or near a band edge \cite{AGT2017PRL}. Importantly, as $\eta$ is increased the  system dynamics requires keeping track of field correlation functions of increasing order.  {We note that the non-Markovianity in this case arises explicitly due to retarded backaction effects, despite having a flat spectral density for the bath.}

 Experimental observations of these effects could be realized across a number of platforms, including quantum dots in photonic waveguides \cite{Kim2018}, atoms near optical nano-fibers \cite{Solano2017,Johnson2019,Kato2019}, and superconducting qubits coupled by coplanar waveguides \cite{vanloo2013,Mirhosseini2018}. Table\,I in the SM \cite{ SM} summarizes  experimental parameters accessible so far.{ For a system of atoms coupled to nanofibers,  values of $\eta\sim 1 $ have already been realized \cite{Kato2019}}. Given the rapid experimental progress in all these platforms, the retarded collective effects studied here can  become relevant in a near future. 
 
{With the emerging possibility of preparing collective dipoles subject to  internal retardation effects and observing their associated complex dynamics  in sharp contrast with the more familiar case of emitters confined in sub-wavelength regions, our work adds a new intricacy that has been little explored in the past.} Given that the enhancement in the retarded collective decay of two emitters relies on a pairwise time-delayed feedback, it will be interesting to determine the scaling of these effects with the number of emitters. We also note that similar dynamics can arise in a system of linear oscillators \cite{JTH2015},  indicating that such collective retarded dissipation should  be observable in classical systems as well.  It would be then interesting to extend the present dynamics from the single-excitation case considered here to multiple excitations, where one can observe genuinely quantum non-Markovian effects, {such as the phenomenon of superfluorescence \cite{Schuurmans} with retardation, where all the emitters decay collectively from a fully excited state}.

%\blue{The effects discussed in this paper demonstrate that dipole-dipole interactions between emitters are modified when they are separated by long-distances. We expect these modifications to become important in scalable quantum networks, distributed quantum sensing and memory schemes. }

\textit{Acknowledgements.}---P.S. and K.S. are grateful to Luis A. Orozco and William D. Philips for initiating the ideas for this work. K.S. would like to thank Alejandro Gonz\'{a}lez-Tudela, Daniel Malz, Darrick E. Chang, and Hyok Sang Han for fruitful discussions, and  Guiseppe Calaj\'{o} for his BIC contribution to the manuscript.

\clearpage
\onecolumngrid
\begin{center}

\newcommand{\beginsupplement}{%
        \setcounter{table}{0}
        \renewcommand{\thetable}{S\arabic{table}}%
        \setcounter{figure}{0}
        \renewcommand{\thefigure}{S\arabic{figure}}%
     }
%\beginsupplement

\textbf{\large Supplemental Material}
\end{center}

\newcommand{\beginsupplement}{%
        \setcounter{table}{0}
        \renewcommand{\thetable}{S\arabic{table}}%
        \setcounter{figure}{0}
        \renewcommand{\thefigure}{S\arabic{figure}}%
     }
%\renewcommand{\thefigure}{S\arabic{figure}}

%%%%%%%%%%% Merge with supplemental materials %%%%%%%%%%
%%%%%%%%%% Prefix a "S" to all equations, figures, tables and reset the counter %%%%%%%%%%
%\setcounter{equation}{0}
%\setcounter{figure}{}
\setcounter{figure}{0}
\setcounter{table}{0}   
\setcounter{page}{1}
\makeatletter
\renewcommand{\theequation}{S\arabic{equation}}
\renewcommand{\thefigure}{S\arabic{figure}}
\renewcommand{\bibnumfmt}[1]{[S#1]}
\renewcommand{\citenumfont}[1]{S#1}
\vspace{0.8 in}

\newcommand{\D}{\Delta}
\newcommand{\tD}{\tilde{\Delta}}
\newcommand{\K}{K_{PP}}
\newcommand{\bn}{\bar{n}_P}
\newcommand{\G}{\Gamma}
\newcommand{\LH}{\underset{L}{H}}
\newcommand{\HL}{\underset{H}{L}}

%\renewcommand{\thefigure}{S\arabic{figure}}
%%%%%%%%%% Prefix a "S" to all equations, figures, tables and reset the counter %%%%%%%%%%
\vspace{-2 cm}

\section{Solution to the dynamics using Lambert $W$-functions}

From taking an inverse Laplace transform of equations (4) and (5) in the main text,  we can write the time dependent atomic excitation  amplitudes as
\eqn{
c_{\substack{\mr{sup}\\\mr{sup}}}(t) &= \frac{1}{2\pi i } \int_{-i\infty +\epsilon}^{+i\infty + \epsilon} \frac{\dd \tilde s }{\sqrt{2}} \frac{e^{\tilde s \gamma t}}{\tilde s+ \frac{1}{2} \pm \frac{1}{2} \beta  e^{ - \eta\tilde s}}\\
\label{cpmt}
& = \frac{1}{2\pi i } \int_{-\infty -i\epsilon}^{+\infty -i \epsilon} \frac{\dd\tilde  z }{\sqrt{2}} \frac{e^{i\tilde z \gamma t}}{\tilde z-\frac{i}{2}  \mp\frac{i}{2}\beta  e^{ - i\eta \tilde z}}.
}
The pole of the denominator is determined by the characteristic equation
\eqn{
&\tilde z-\frac{i}{2}  \mp\frac{i}{2}\beta    e^{-i\eta \tilde z}=0
\implies\bkt{i\eta\bar z} e^{i\eta \bar z} = \mp\frac{\eta}{2}\beta  e^{\eta/2}
\implies\tilde  z_n^{(\pm)} = \frac{i}{2}\sbkt{ 1- \frac{W_n\bkt{\mp \beta  \frac{\eta}{2}e^{\eta/2} } }{\eta/2}},
}
where we have introduced $\bar z \equiv \tilde z-i  /2$ in the intermediate step. Here $W(z)$ is the Lambert $W$-function, or more precisely a set of functions $W_n(z)$ comprising the $n$ branches of the inverse relation of the function $f(z) = ze^z$, where $z$ is a complex number. In other words $ z=f^{-1}\left(ze^{z}\right) \equiv W\left(ze^{z}\right)$, $W(z) = f^{-1}(z) $, and $W_n(x)$ is its $n$-th branch~\cite{Corless96}. 

For a given value of $\beta $ there is a critical distance between the atoms defined by  $\beta \frac{\eta_c}{2} e^{\eta_c/2}  = 1/e $, or $\eta_c  =2W_0\bkt{1/e \beta},$ such that for $\eta > \eta_c $ the dominant pole contributions from $\tilde z_{0}$ acquire a real component, leading to oscillatory dynamics. We can now write the Laurent series expansion of the denominator of the integrand in \eqref{cpmt}
\eqn{\frac{1}{\tilde z - \frac{i}{2} \mp\frac{i}{2} \beta e^{- i \eta \tilde  z}}  &= \sum_{n\in \mathbb{Z}} \frac{\alpha_n^{(\pm)}}{\tilde z - \tilde z_n^{(\pm)}},}
such that 
\eqn{\alpha_n^{(\pm)} = \lim_{\tilde z\rightarrow{\tilde z_n^{(\pm)}}} \frac{\tilde z-\tilde z_n^{(\pm)} }{\tilde z - \frac{i}{2} \mp\frac{i}{2}\beta  e^{ - i \eta \tilde  z}} 
& = \frac{1}{1+W_n\bkt{\mp\beta\frac{\eta}{2}  e^{\eta/2} } },
}
where we have used the property of the $W$-function that $W(z_0) e^{W(z_0)} = z_0 $ \footnote{It can be seen from the definition of the $W$-function that given $W(z e^z) =z$, if we substitute $y \equiv z e^z$ in this definition, one obtains $W(y) = z$. Substituting $W(y) = z$ back in $y \equiv z e^z$ yields $y = W(y) e^{W(y) }$.}. This gives 

\eqn{\frac{1}{\tilde z - \frac{i}{2} \mp\frac{i}{2} \beta e^{- i \eta \tilde  z}}  & =  \sum_{n\in \mathbb{Z}} \frac{1}{\cbkt{1+W_n\bkt{ \mp\frac{\eta}{2}\beta e^{\eta/2}}} \bkt{\tilde z - \tilde z_n^{(\pm)}}}, 
}
such that we can write the inverse Laplace transform of \eqref{cpmt} as
\eqn{c_{\substack{\mr{sup}\\\mr{sub}}}(t) = \frac{ 1}{\sqrt{2}}\sum_{n \in \mathbb{Z}} \alpha_n^{(\pm)} e^{ -{\gamma_n^{(\pm)}}  t/2} ,
}
where we have defined \eqn{\alpha_n^{(\pm)} &\equiv \frac{1}{1+ {W_{n} \bkt{\mp\frac{\eta}{2}\beta  e^{\eta/2 } } }}\\
\gamma_ n^{(\pm)} &\equiv \gamma \sbkt{1-\frac{W_{n} \bkt{\mp\frac{\eta}{2}\beta  e^{\eta/2 } } }{\eta/2}},}
as given in the main text.

We remark that the Lambert $W$-function dependence of the dynamics on the separation is a characteristic of time-delayed systems with self-consistent retarded backaction \cite{Corless96} that has been studied in detail in control theory literature and in a variety of other physical problems  ranging from instrument design \cite{Inst} to the AdS/CFT correspondence \cite{Ads}.

We also note that while the real part of $\gamma^{(\pm)}$ leads to non-Markovian effects in collective spontaneous emission, the imaginary part can similarly lead to  delay-induced collective effects in the van der Waals shifts between the emitters \cite{Sinha18, Fuchs18}.

\begin{figure}
    \centering
    \subfloat[]{\includegraphics[width = 4 in]{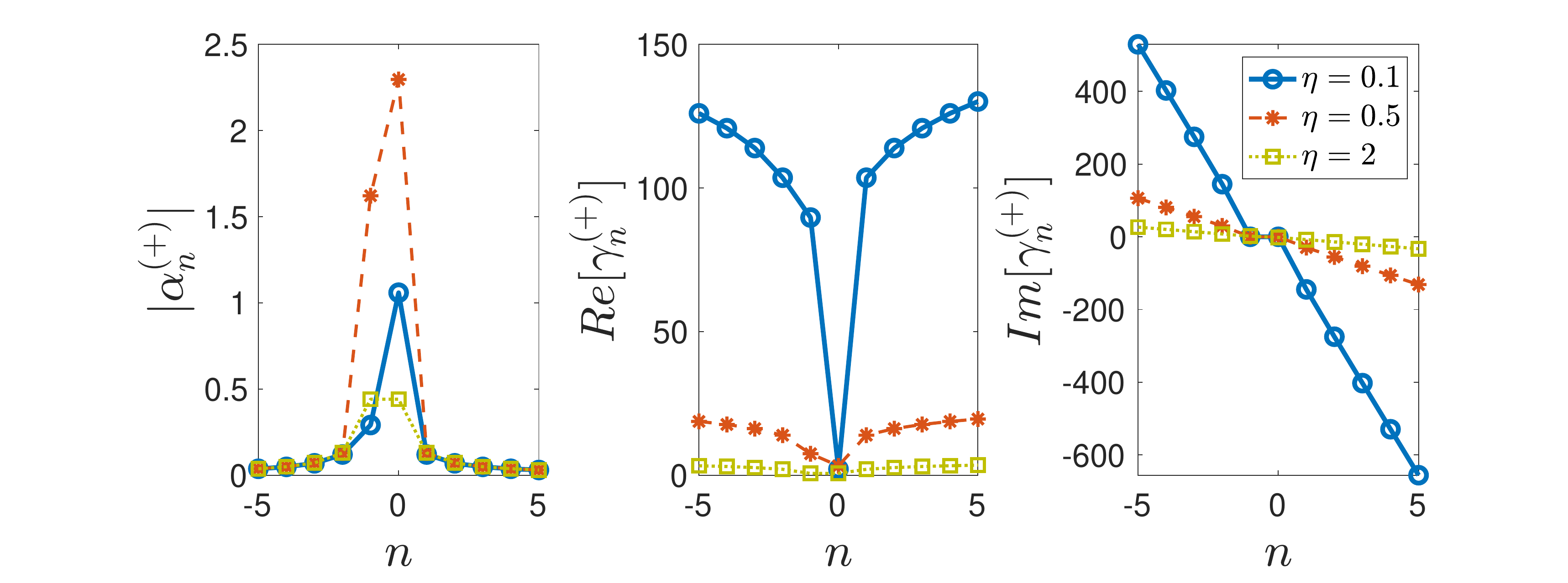}}\\
       \subfloat[]{\includegraphics[width = 4 in]{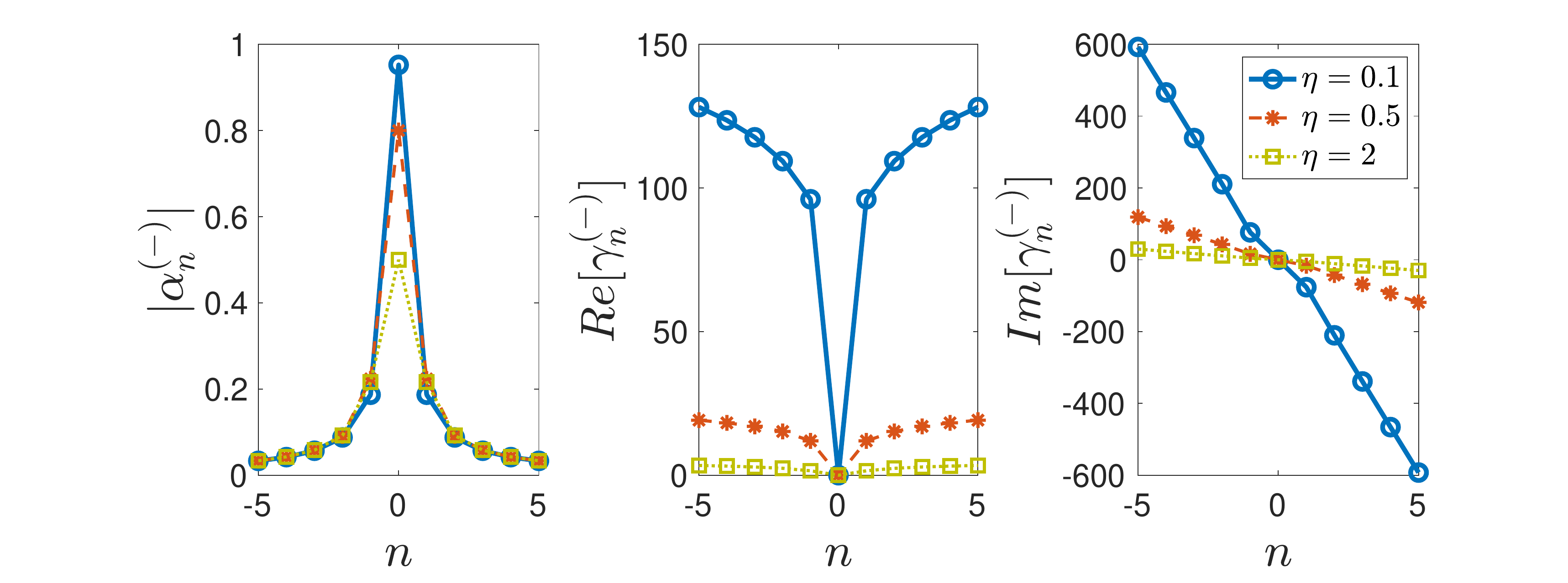}}
    \caption{Coefficients $\alpha_n^{(\pm )}$ and $\gamma_n^{(\pm) }$ for $\beta = 1$ and $\gamma =1$ for different values of $n $ and $\eta$ parameter.}
    \label{FigS1}
\end{figure}
\section{Alternative solution in terms of wavepacket oscillations}
Consider Eqs.\,(4) and (5) from the main text,  noting that $\abs{\bkt{\tilde s+1/2}/\bkt{\beta/2 e^{ -\eta \tilde s}}}<1$ such that the denominator in Eqs.\,(4) and (5) in the main text  be expanded as the following series
\eqn{
c_{\substack{\mr{sup}\\\mr{sub}}}(t) \approx \frac{1}{2\pi i } \int_{-i\infty +\epsilon}^{+i\infty + \epsilon} \frac{\dd \tilde{s} }{\sqrt{2}} \frac{e^{\gamma\tilde{s}t}}{\tilde {s}+1/2}\sbkt{\sum_{j = 0 }^\infty  \frac{(\pm\beta)^j e^{ - j\eta \tilde s}}{\bkt{\tilde s + 1/2}^j }}.
}
The dynamics in this limit has been previously studied \cite{Milonni74, Arecchi70, Dorner02}, leading to the atomic excitation amplitudes.

Taking the inverse Laplace transform in the above equation, we obtain
\eqn{\label{peters}
c_{\substack{\mr{sup}\\\mr{sub}}}(t) \approx\frac{1}{\sqrt{2}} \sum_{j = 0 }^\infty&\frac{\bkt{\mp\beta}^j}{j!} \bkt{ \frac{\gamma t-j\eta}{2} }^je^{ - \bkt{ \gamma t - j\eta }/2} \Theta \bkt{ \gamma t - j\eta},
}
 where the index $j$ physically corresponds to the number of round trips of the photon wavepacket between the two emitters. The multiple reflections of the photon wavepacket modify the time evolution of the spontaneous decay, making it non-exponential, as given in \cite{Milonni74}. It is helpful to consider this physical picture and expansion in the large $\eta$ limit, as it allows one to see how the wavepackets bounce between the atoms.
 However, the solution in terms of $W$-functions allows one to understand the collective dynamics of the emitters more effectively, allowing one to -- (1) determine the critical separation $\eta_c$ up to which the emitters decay monotonically, as opposed to exhibiting oscillatory dynamics, (2) evaluate the instantaneous spontaneous emission rate for the atomic decay after the onset of collective dynamics, and (3) calculate the probabilities of atomic excitation in the steady state when a BIC state is formed.

\section{Field Intensity}
One can write the dynamics of the field amplitudes from equations (1) and (2) in the main text as follows 
\eqn{
c_a\bkt{\omega, t} &= -i\sqrt{\frac{\gamma\beta}{4\pi}} \int_0 ^t \dd\tau\, \sum_{m = 1, 2} c_m\bkt{\tau} e^{-i \omega x_m/v_g} e^{i \bkt{\omega - \omega_0 }\tau }\\
c_b\bkt{\omega, t} &= -i\sqrt{\frac{\gamma\beta}{4\pi}} \int_0 ^t \dd\tau\, \sum_{m = 1, 2}  c_m\bkt{\tau} e^{i \omega x_m/v_g} e^{i \bkt{\omega - \omega_0 }\tau },
}
we note that for an initial  superradiant (subradiant) state, the fields emitted into the left and right going modes have same (opposite) phases such that
\eqn{\label{cab}
c_a^{\mr{sup}}\bkt{\omega, t} &= c_b^{\mr{sup}}\bkt{\omega, t} = -i\sqrt{\frac{\gamma \beta}{2\pi}} \sum_{n\in\mathbb{Z}} \alpha_n ^{(+)}\cos\bkt{\frac{kd}{2}}\frac{ e^{\sbkt{i \bkt{\omega - \omega_0 } -\gamma_n ^{(+)}/2}t} - 1}{i \bkt{\omega - \omega_0 } -\gamma_n ^{(+)}/2}\\
c_a^{\mr{sub}}\bkt{\omega, t} &= -c_b^{\mr{sub}}\bkt{\omega, t}= -i\sqrt{\frac{\gamma\beta}{2\pi}} \sum_{n\in\mathbb{Z}} \alpha_n ^{(-)}\sin\bkt{\frac{kd}{2}}\frac{ e^{\sbkt{i \bkt{\omega - \omega_0 } -\gamma_n ^{(-)}/2}t} - 1}{i \bkt{\omega - \omega_0 } -\gamma_n ^{(-)}/2},
}
where we have used  Equation (9) from the main text for the emitter dynamics. The superscripts refer to the initial state of the emitters being in a symmetric (sup) or anti-symmetric (sub) superposition.

The intensity of the field emitted by the atoms as a function of position and time can be evaluated as $I\bkt{x,t} = \frac{\epsilon_0 c} {2}\avg{\Psi\bkt{t}\vert\hat{E}^\dagger\bkt{x,t} \hat E\bkt{x,t}\vert\Psi\bkt{t}} $, where $\hat{E}\bkt{x,t} = \int_0^\infty  \dd k\,\mc{E}_k \sbkt{\hat{a} \bkt{k}e^{i k x} + \hat{b} \bkt{k}e^{-i k x}}e^{i \omega t}$ is the electric field operator at position $x$ and time $t$. We assume $\mc{E}_k\approx\mc{E}_{k_0}$ to be constant for all $k$. More explicitly, for the superradiant state we obtain 
\eqn{
I_\mr{sup}(x,t)/I_0  =&  \bra{\Psi\bkt{t}}\sbkt{\int \dd k_1  \cbkt{\hat{a}^\dagger \bkt{k_1}e^{-i k_1 x} + \hat{b}^\dagger \bkt{k_1}e^{i k_1 x}}e^{-i \omega_1 t}\right.\non\\
&\left.\int \dd k_2\cbkt{\hat{a} \bkt{k_2}e^{i k_2 x} + \hat{b} \bkt{k_2}e^{-i k_2 x}}e^{i \omega_2 t}} \ket{\Psi\bkt{t}}\\
 = & \int \dd k_1  \int \dd k_2  \sbkt{ e^{-i \bkt{ k_1 - k_2} x} c_a^\ast \bkt{ \omega_1, t} c_a\bkt{ \omega_2, t} + e^{i \bkt{ k_1 - k_2} x} c_b^\ast \bkt{ \omega_1, t} c_b\bkt{ \omega_2, t} \right.\non\\
 &\left. + e^{i \bkt{ k_1 + k_2} x} c_b^\ast \bkt{ \omega_1, t} c_a\bkt{ \omega_2, t} + e^{-i \bkt{ k_1 + k_2} x} c_a^\ast \bkt{ \omega_1, t} c_b\bkt{ \omega_2, t}}e^{-i \bkt{\omega_1 -\omega_2}t}\\
 = &\int\dd k_1 \int\dd k_2 \,c_a^\ast \bkt{ \omega_1, t} c_a\bkt{ \omega_2, t}  \cos\bkt{k_1x} \cos\bkt{k_2x} e^{-i \bkt{\omega_1 -\omega_2}t}\\
 = &\abs{\int\dd k\,c_a \bkt{ \omega, t}  \cos\bkt{kx}e^{-i \omega t}}^2,
}
where we have used the fact that the field amplitudes $c_a (\omega, t ) = c_b (\omega, t)$ for an initial  superradiant state, which also leads to a symmetric distribution around $x=0$. The normalization constant $I_0 = \frac{\epsilon_0 c \mc{E}_{k_0}^2}{2}$.
We thus obtain the intensity as
\eqn{ I_{\mr{sup}}(x,t)/I_0=& \frac{2\gamma\beta} { \pi }\abs{ \sum_n \alpha_n^{(+)} \int _0 ^\infty \dd k \cos\bkt{kx }\cos\bkt{kd/2 } \cbkt{ \frac{ e^{-\bkt{i \omega_0+\gamma_n^{(+)}/2} t} - e^{-i\omega t}}{ i \bkt{\omega - \omega_0 } - \gamma_n ^{(+) }/2}} }^2\\
= & \frac{\gamma\beta\pi }{2 } \abs{ \sum_{n\in \mathbb{Z}} \alpha_n^{(+)}\sbkt{ \underbrace{\cbkt{\Theta\bkt{t - \tau_1} - \Theta\bkt{- \tau_1 }}e^{-\bkt{i \omega_0 + \frac{\gamma_n^{(+)}}{2}}\bkt{t - \tau_1}}}_{\text{1a}}+ \underbrace{ \cbkt{\Theta\bkt{t + \tau_1} - \Theta\bkt{ \tau_1 }}e^{-\bkt{i \omega_0 +\frac{ \gamma_n^{(+)}}{2}}\bkt{t + \tau_1}}}_{\text{1b}} \right.\right.\non\\
&\left.\left.+\underbrace{\cbkt{\Theta\bkt{t - \tau_2} - \Theta\bkt{- \tau_2 }}e^{-\bkt{i \omega_0 + \frac{\gamma_n^{(+)}}{2}}\bkt{t - \tau_2}}}_{\text{2a}}+ \underbrace{\cbkt{\Theta\bkt{t + \tau_2} - \Theta\bkt{ \tau_2 }}e^{-\bkt{i \omega_0 +\frac{ \gamma_n^{(+)}}{2}}\bkt{t + \tau_2} }}_{\text{2b}} } }^2 ,}
where $\tau_{\substack{1\\2}}\equiv x\pm d/2$, the terms $n$a and $n$b correspond to the fields emitted by atom $n = 1,2$ into the left and right going modes, a and b, respectively. The terms $n$a and $n$b together make up the  light cones emanating from the $n$th emitter. It can be seen that the two light cones interfere constructively with each other going outwards from the atoms and contribute to a beyond superradiant burst, as can be seen from Fig.\,4 of the main text.

Similarly for the subradiant state 
\eqn{I_\mr{sub}(x,t)/I_0  = & \frac{2\gamma\beta}{ \pi } \abs{ \sum_n \alpha_n^{(-)} \int _0 ^\infty \dd k \sin\bkt{kx }\sin\bkt{kd/2 } \cbkt{ \frac{ e^{-\bkt{i \omega_0+\gamma_n^{(-)}/2} t} - e^{-i\omega t}}{ i \bkt{\omega - \omega_0 } - \gamma_n ^{(-) }/2}} }^2}
We note that as an important difference from the superradiant case, for $n = 0 $, $\gamma_n^{(-)}=0$, which leads the integral $\int_0 ^\infty \dd k \frac{e^{-i\omega_0 t } - e^{-i\omega t}}{ i\bkt{\omega - \omega_0 }}$ to diverge due to a pole contribution on the real axis (which corresponds to a divergent self-energy term otherwise). We therefore take only the principal value of the integral to obtain
\eqn{ I_\mr{sub}(x,t)/I_0
= & \frac{\gamma\beta\pi }{2 } \abs{ \sum_{n\in \mathbb{Z}} \alpha_n^{(-)}\sbkt{ \underbrace{\cbkt{\Theta\bkt{t - \tau_1} - \Theta\bkt{- \tau_1 }}e^{-\bkt{i \omega_0 + \frac{\gamma_n^{(-)}}{2}}\bkt{t - \tau_1}}}_{\text{1a}}+ \underbrace{\cbkt{\Theta\bkt{t + \tau_{1}} - \Theta\bkt{ \tau_1 }}e^{-\bkt{i \omega_0 +\frac{ \gamma_n^{(-)}}{2}}\bkt{t + \tau_1}}}_{\text {1b}} \right.\right.\non\\
&\left.\left.-\underbrace{\cbkt{\Theta\bkt{t - \tau_{2}} - \Theta\bkt{- \tau_2 }}e^{-\bkt{i \omega_0 + \frac{\gamma_n^{(-)}}{2}}\bkt{t - \tau_2}}}_{\text{2a}}-\underbrace{ \cbkt{\Theta\bkt{t + \tau_2} - \Theta\bkt{ \tau_2 }}e^{-\bkt{i \omega_0 +\frac{ \gamma_n^{(-)}}{2}}\bkt{t + \tau_2}}} _{\text {2b}}}}^2.}

Again, similar to the superradiant case, the four terms correspond to fields emitted by the two atoms into the left and right propagating field modes, which in this case interfere destructively with each other outside of the atomic cavity, and form a standing wave inside of the atomic cavity, as illustrated in Fig.4 in the main text.

 \section{Cooperativity}
We remark that the cooperativity for the system  can be calculated as 
$\mc{C} = \gamma_\mr{in}/\gamma_{3D}$ as discussed in the main text. The total emission into the waveguide can be calculated as 
\eqn{
\gamma_\mr{in} = \lim_{t\rightarrow{}\infty}\int_0^\infty \dd\omega \,\sbkt{ \abs{c_a(\omega, t)}^2 + \abs{c_b(\omega, t)}^2}.
}
Substituting \eqref{cab} in the above this can be simplified to Eq.\,(10) in the main text.

\section{Non-Markovianity measure via coherence}

We calculate the \textit{coherence measure of non-Markovianity} as defined in \cite{Chanda16}. We first calculate the $l_1$-norm of coherence as

\eqn{
\mc{C}\bkt{\rho\bkt{t}} = \sum _{\substack{i,j\\{i\neq j}}} \abs{\rho_{ij}\bkt{t}},
}
where $\rho(t)$ refers to the system density matrix corresponding to  the two emitters given by 
\eqn{\rho(t)  =\bkt{ \begin{array}{cccc}
     \rho_{ee,ee}&\rho_{ee,eg} &\rho_{ee,ge} &\rho_{ee,gg}  \\
     \rho_{eg,ee}&\rho_{eg,eg} &\rho_{eg,ge} &\rho_{eg,gg}  \\
     \rho_{ge,ee}&\rho_{ge,eg} &\rho_{ge,ge} &\rho_{ge,gg}  \\
     \rho_{gg,ee}&\rho_{gg,eg} &\rho_{gg,ge} &\rho_{gg,gg}  
\end{array}}
}
where, for an intial superradiant state
\eqn{
\rho^\mr{sup}_{ee,ee}&=\rho^\mr{sup}_{ee,eg} =\rho^\mr{sup}_{ee,ge} =\rho^\mr{sup}_{ee,gg}=\rho^\mr{sup}_{eg,gg} =\rho^\mr{sup}_{ge,gg} =0\\
\rho^\mr{sup}_{eg,eg} &= \rho^\mr{sup}_{ge,ge}=\rho^\mr{sup}_{eg,ge} =\rho^\mr{sup}_{ge,eg} = \abs{c_\mr{sup}(t) }^2\\
\rho^\mr{sup}_{gg,gg} &= 1 - 2\abs{c_\mr{sup}(t) }^2 }
For an initial subradiant state
\eqn{
\rho^\mr{sub}_{ee,ee}&=\rho^\mr{sub}_{ee,eg} =\rho^\mr{sub}_{ee,ge} =\rho^\mr{sub}_{ee,gg}=
\rho^\mr{sub}_{eg,gg} =\rho^\mr{sub}_{ge,gg} =0\\
\rho^\mr{sub}_{eg,eg} &= \rho^\mr{sub}_{ge,ge}=-\rho^\mr{sub}_{eg,ge} =-\rho^\mr{sub}_{ge,eg} = \abs{c_\mr{sub}(t) }^2\\
\rho^\mr{sub}_{gg,gg} &= 1 - 2\abs{c_\mr{sub}(t) }^2
}
This yields the $l_1$-norm of the coherence for the super- and sub-radiant states as 
\eqn{
\mc{C}_{\substack{\mr{sup}\\\mr{sub}}} &= \pm2 \abs{c_{\substack{\mr{sup}\\\mr{sub}}}(t) }^2 = \pm\abs{\sum_{n \in \mathbb{Z}}  \alpha_n^{(\pm)} e^{ -\gamma_n^{(\pm)}  t/2}}^2 
}

Further considering a general initial state in the single-excitation subspace \eqn{\ket{\Psi _{\theta, \phi }} \equiv \cos \theta \ket{\Psi_\mr{sup}} + e^{i \phi}\sin \theta  \ket{\Psi_\mr{sub}},
}
we find the time evolved state as
\eqn{
\ket{\Psi_{\theta, \phi }(t) } \equiv \cos \theta \ket{\Psi_\mr{sup}(t)} + e^{i \phi}\sin \theta  \ket{\Psi_\mr{sub}(t)}.
}
The $l_1$-norm of the coherence for the above state is given as
\eqn{\mc{C}_{\theta, \phi} = 2 \sbkt{\cos^2\theta \abs{c_\mr{sup}(t)}^2 - \sin^2\theta \abs{c_\mr{sub}(t)}^2}.
}

For a given  initial  state, one can further define the coherence measure of non-Markovianity as
\eqn{
\mc{N}= \mr{max}_{\rho(0)\in \cbkt{\ket{\Psi_{\theta,\phi}}} } \int_{\der {  \mc{C}\bkt{\rho\bkt{t}} }{t}>0} \dd t \der {  \mc{C}\bkt{\rho\bkt{t}} }{t}.
}
In the above expression,  we calculate the non-Markovianity measure for a restricted initial state space of the system \cite{Chanda16}, considering the initial state space of single emitter excitation. Since the integral is defined over only positive values of $d\mc{C}/dt$, a non-zero value of $\mc{N}$ immediately implies that the system is non-Markovian.

We can also  obtain a non-Markovianity measure for the super- and sub-radiant states as
\eqn{
\mc{N}_{\substack{\mr{sup}\\\mr{sub}}} = \int_{\mc{C}'_{\substack{\mr{sup}\\\mr{sub}}}>0}\dd t \bkt{\abs{\sum_{n \in \mathbb{Z}}  \alpha_n^{(\pm)} e^{ -\gamma_n^{(\pm)}  t/2}}}\bkt{\abs{\sum_{m \in \mathbb{Z}}  \alpha_m^{(\pm)} \gamma_m^{(\pm)} e^{ -\gamma_m^{(\pm)}  t/2}}}.
}

We now optimize the coherence measure of non-Markovianity over the initial state parameters to obtain $\mc{N} = \mc{N}_\mr{sub}$. This shows that according to the coherence measure of non-Markovianity, an initially subradiant states remains the most non-Markovian throughout the evolution.
\begin{figure}[htb]
    \centering
    \includegraphics[width = 3.5 in]{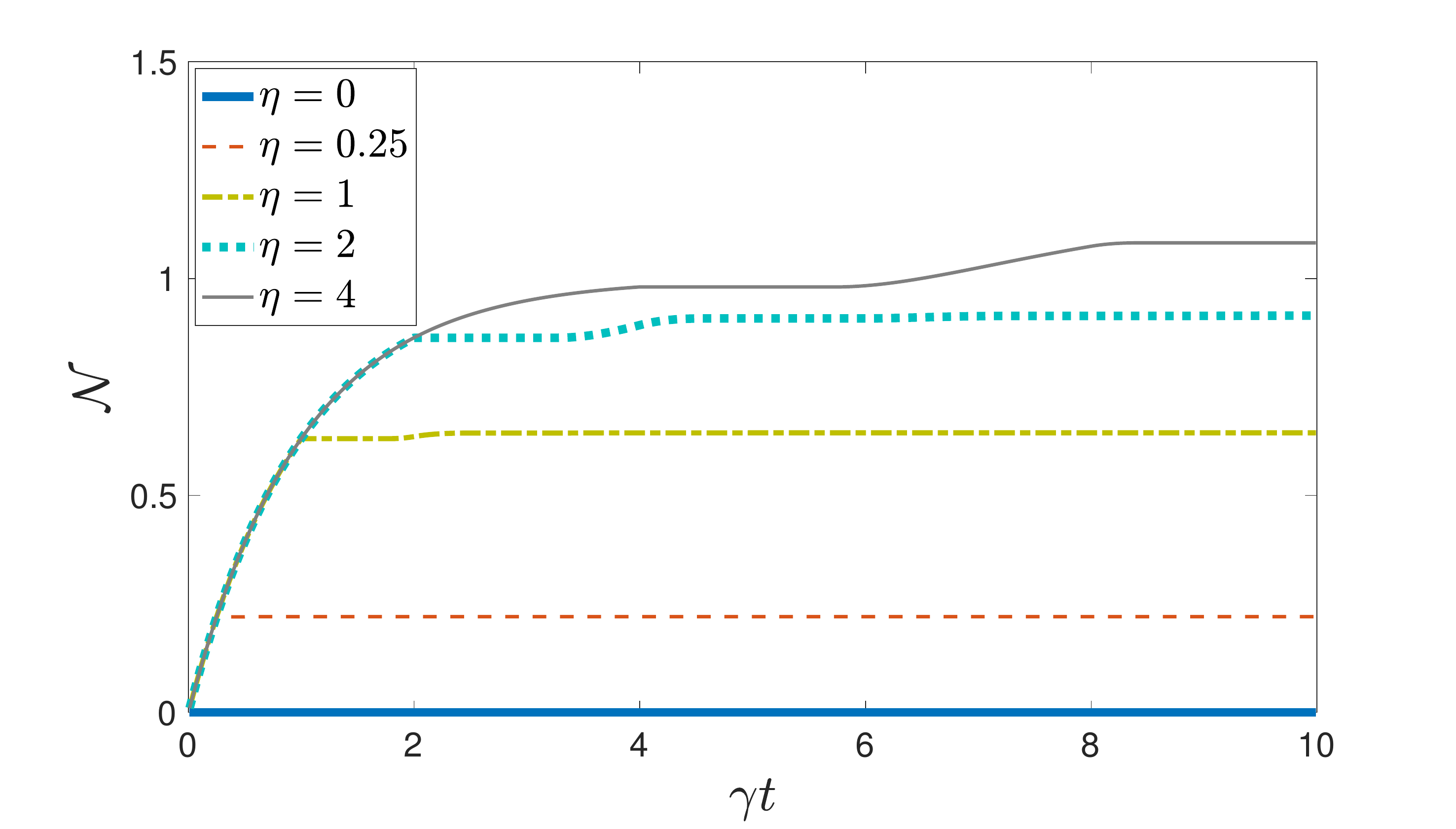}
    \caption{Coherence measure of non-Markovianity for  the system of  emitters initially prepared in the single excitation subspace  as a function of time, for different values of the delay. }
    \label{FignonM}
\end{figure}

\section{Correlation between the emitters and the field}

We calculate the correlations between the emitters and the EM field via the linear entropy of the reduced system density matrix defined as \cite{Bose2000}

\begin{figure}[htb]
    \centering
    \includegraphics[width = 3.5 in]{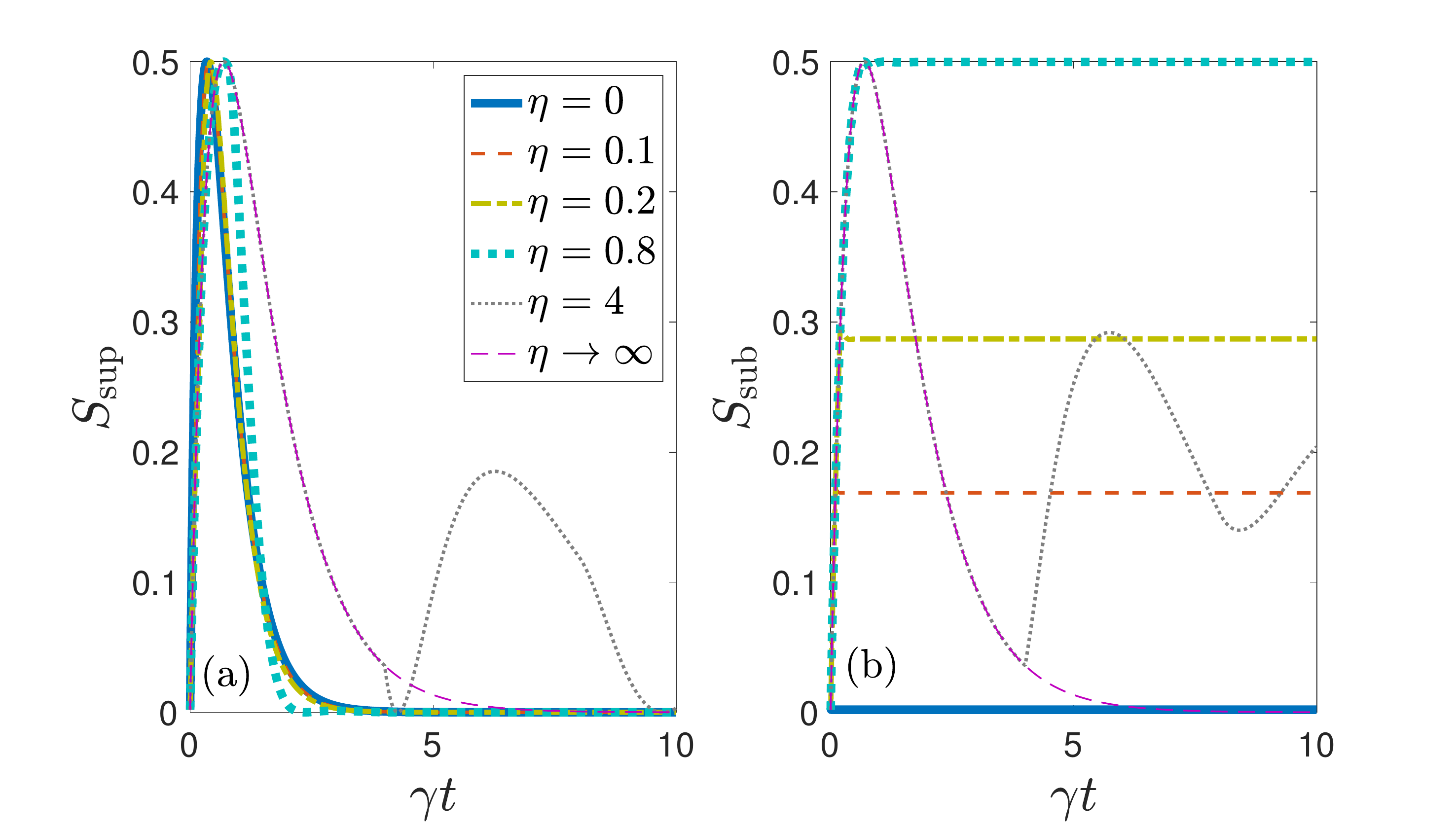}
    \caption{Linear entropy of entanglement between the emitters and the EM field as a function of time for initial (a) super- and (b) sub-radiant states of the emitters with different delays. We note from the dotted line in  (b)  that the emitter-field entanglement is optimized for a specific value of the delay $\eta\approx0.8$ for an initial subradiant state of the emitters. This corresponds to a BIC.}
    \label{FigSL}
\end{figure}

\eqn{
S = 1- \mr{Tr} \sbkt{ \rho ^2}.
}
Evaluating the above for a super-and sub-radiant state
\eqn{
S _{\substack{\mr{sup}\\\mr{sub}}} &= 4\abs{c_ {\substack{\mr{sup}\\\mr{sub}}}}^2\bkt{ 1 - 2\abs{c_ {\substack{\mr{sup}\\\mr{sub}}}}^2}\\
 &= 2 \abs{\sum_{m \in \mathbb{Z}}  \alpha_m^{(\pm)} \gamma_m^{(\pm)} e^{ -\gamma_m^{(\pm)}  t/2}}^2\bkt{1 - \abs{\sum_{m \in \mathbb{Z}}  \alpha_m^{(\pm)} \gamma_m^{(\pm)} e^{ -\gamma_m^{(\pm)}  t/2}}^2 }.
}
We plot the above linear entropy of entanglement in Fig.\,\ref{FigSL}.

\section{Possible implementations}
Table\,\ref{table} we lists some potential systems that study the system of two emitters coupled to waveguides experimentally, and the parameter values that have been realized in these platforms. 
\begin{center}
\begin{table*}[htb]
\label{table}
\begin{tabular}{|c|c|c|c|c|c|c|}

\hline
Emitter   & $\omega_0/2\pi $(THz) & $\gamma_0/2\pi $(MHz) & $\gamma_{1\text{D}}/\gamma$ & $v_g/c$ & d (m)              & $\eta$  \\ 
\hline
Atoms \cite{Kato2019}     & 380                   & 6                     & 0.1                           & 0.7-0.1 & 2                  & $\sim 1$    \\
Quantum dots (QD)   \cite{Kim2018}     & 230                   & $17\times10^{3}$      & 0.7                           & 0.08    & $15\times10^{-6}$  & $\sim 0.07$ \\
Superconducting circuits  (SC) \cite{vanloo2013}     & $6\times10^{-3}$      & 20                    & 0.9                           & 0.5     & $18\times 10^{-3}$ & $\sim 0.02$  \\
% Molecules \cite{Turschmann2017} & 380  &      45 &     0.07        &     0.6    &        $50 \times 10^{-6}$    &       $10^{-5}$   \\
\hline
\end{tabular}
\caption{List of experimental parameters achieved up to date in different platforms.}
\end{table*}
\end{center}


\vspace{-0.5 cm}


\begin{thebibliography}{92}
% applications, long range quantum networks etc {Kimble08, Schoelkopf08, Pichler17, Pichler16}
\bibitem{Kimble08} H. J. Kimble, The quantum internet, \href{https://doi.org/10.1038/nature07127}{ Nature {\bf453}, 1023 (2008)}.
    \bibitem{Schoelkopf08} R. J. Schoelkopf and S. M. Girvin, Wiring up quantum systems, \href{https://doi.org/10.1038/451664a}{Nature {\bf 451}, 664 (2008)}.
    \bibitem{Pichler17}H. Pichler, S. Choi, P. Zoller, and M. D. Lukin,  Universal photonic quantum computation via time-delayed feedback, \href{https://doi.org/10.1073/pnas.1711003114
    }{ Proc. Natl. Acad. Sci.  {\bf114}, 11362 (2017)}.
    \bibitem{Pichler16} H. Pichler and P. Zoller, Photonic Circuits with Time Delays and Quantum Feedback, \href{https://doi.org/10.1103/PhysRevLett.116.093601}{Phys. Rev. Lett. {\bf116}, 093601 (2016)}.
    \bibitem{Ge2018} W. Ge, K. Jacobs, Z. Eldredge, A. V. Gorshkov, and M. Foss-Feig, Distributed Quantum Metrology with Linear Networks and Separable Inputs \href{https://doi.org/10.1103/PhysRevLett.121.043604}{ Phys. Rev. Lett. 121, 043604 (2018)}.
    % Coherent time delayed feedback, non-Markovian dynamics
    \bibitem{Whalen17}S. J. Whalen, A. L. Grimsmo, and H. J. Carmichael, Open quantum systems with delayed coherent feedback, \href{https://doi.org/10.1088/2058-9565/aa8331}{Quantum Sci. Technol. {\bf2}, 044008 (2017)}.
    \bibitem{Grimsmo15} A. L. Grimsmo, Time-Delayed Quantum Feedback Control, \href{https://doi.org/10.1103/PhysRevLett.115.060402}{ Phys. Rev. Lett. {\bf115}, 060402 (2015)}.
    % nonMarkovian atom field interactions {Breuer16, BPbook, deVega17,HPZ92, HPZ93, Vasile14, Groblacher15  Fleming12}
    \bibitem{BPbook} H.-P. Breuer, and F. Petruccione, \textit{Theory of open quantum systems} (Oxford University Press, New York, 2002).
    
    \bibitem{Breuer16}H. P. Breuer, E. M. Laine, J. Piilo, and B. Vacchini, Colloquium: Non-Markovian dynamics in open quantum systems,  \href{https://doi.org/10.1103/RevModPhys.88.021002}{ Rev. Mod. Phys. {\bf88}, 021002 (2016)}.
    \bibitem{deVega17} I. de Vega, and D. Alonso, Dynamics of non-Markovian open quantum systems, \href{https://doi.org/10.1103/RevModPhys.89.015001}{ Rev. Mod. Phys. {\bf89}, 015001 (2017)}.
    
    \bibitem{Fleming12} C. Fleming, and B. L. Hu, Non-Markovian dynamics of open quantum systems: Stochastic equations and their perturbative solutions, \href{https://doi.org/10.1016/j.aop.2011.12.006}{Ann. Phys. {\bf 327}, 1238 (2012)}. 

% nonmarkovian spectral density {}
\bibitem{HPZ92} B. L. Hu, J. P. Paz, and Y. Zhang, Quantum Brownian motion in a general environment: Exact master equation with nonlocal dissipation and colored noise, \href{https://doi.org/10.1103/PhysRevD.45.2843}{Phys. Rev. D {\bf45}, 2843 (1992)}.
\bibitem{HPZ93} B. L. Hu, J. P. Paz, and Y. Zhang, Quantum Brownian motion in a general environment. II. Nonlinear coupling and perturbative approach,  \href{https://doi.org/10.1103/PhysRevD.47.1576}{Phys. Rev. D 47, 1576 (1993)}.

\bibitem{Vasile14} R. Vasile, F. Galve, and R. Zambrini,  Spectral origin of non-Markovian open-system dynamics: A finite harmonic model without approximations, \href{https://doi.org/10.1103/PhysRevA.89.022109}{ Phys. Rev. A {\bf89}, 022109 (2014)}.
\bibitem{Groblacher15} S. Gr\"{o}blacher, A. Trubarov, N. Prigge, G. D. Cole, M. Aspelmeyer, and J. Eisert, Observation of non-Markovian micromechanical Brownian motion, \href{https://doi.org/10.1038/ncomms8606}{  Nat. Commun. {\bf6}, 7606 (2015)}.



% \bibitem{Lin09}S.-Y. Lin, and B. L. Hu, Temporal and spatial dependence of quantum entanglement from a field theory perspective, \href{https://doi.org/10.1103/PhysRevD.79.085020}{ Phys. Rev. D {\bf79}, 085020 (2009)}.

\bibitem{JTH2015} J.-T. Hsiang, and B. L. Hu, Distance and coupling dependence of entanglement in the presence of a quantum field,  \href{https://doi.org/10.1103/PhysRevD.92.125026}{Phys. Rev. D {\bf92}, 125026 (2015).}

\bibitem{Giessen96}H. Gie{\ss}en, J. D. Berger, G. Mohs, P. Meystre, and S. F. Yelin, Cavity-modified spontaneous emission: From Rabi oscillations to exponential decay,  \href{https://doi.org/10.1103/PhysRevA.53.2816}{ Phys. Rev. A {\bf53}, 2816 (1996)}.

\bibitem{Dorner02} U. Dorner and P. Zoller, Laser-driven atoms in half-cavities,  \href{https://doi.org/10.1103/PhysRevA.66.023816}{Phys. Rev. A {\bf66}, 023816 (2002)}.
\bibitem{Tufarelli13} T. Tufarelli, F. Ciccarello, and M. S. Kim, Dynamics of spontaneous emission in a single-end photonic waveguide,  \href{https://doi.org/10.1103/PhysRevA.87.013820}{ Phys. Rev. A {\bf87}, 013820 (2013)}.
\bibitem{Tufarelli14} T. Tufarelli, M. S. Kim, and F. Ciccarello, Non-Markovianity of a quantum emitter in front of a mirror, \href{https://doi.org/10.1103/PhysRevA.90.012113}{ Phys. Rev. A {\bf90}, 012113 (2014)}.
\bibitem{Carmele13} A. Carmele, J. Kabuss, F. Schulze, S. Reitzenstein, and A. Knorr, Single Photon Delayed Feedback: A Way to Stabilize Intrinsic Quantum Cavity Electrodynamics,  \href{https://doi.org/10.1103/PhysRevLett.110.013601}{Phys. Rev. Lett. {\bf110}, 013601 (2013)}.
\bibitem{Cook87} R. J. Cook and P. W. Milonni, Quantum theory of an atom near partially reflecting walls,  \href{https://doi.org/10.1103/PhysRevA.35.5081}{Phys. Rev. A {\bf35}, 5081 (1987)}.
\bibitem{Beige02} A. Beige, J. Pachos, and H. Walther, Spontaneous emission of an atom in front of a mirror, \href{https://doi.org/10.1103/PhysRevA.66.063801}{Phys. Rev. A {\bf66}, 063801 (2002)}.

\bibitem{Guimond16} P.-O. Guimond, A. Roulet, H. N. Le, and V. Scarani, Rabi oscillation in a quantum cavity: Markovian and non-Markovian dynamics, \href{https://doi.org/10.1103/PhysRevA.93.023808}{Phys. Rev. A {\bf93}, 023808 (2016).}
\bibitem{Guimond17} P.-O. Guimond, M. Pletyukhov, H. Pichler, and P. Zoller, Delayed coherent quantum feedback from a scattering theory and a matrix product state perspective \href{https://doi.org/10.1088/2058-9565/aa7f03}{ Quantum Sci. Technol. {\bf2}, 044012 (2017)}.

% BIC references {Hsu16, Fong17, Calajo19, Facchi19}
\bibitem{Hsu16} C. W. Hsu, B. Zhen, A. D. Stone, J. D. Joannopoulos, and M. Solja\v{c}i\'{c}, Bound states in the continuum, \href{https://doi.org/10.1038/natrevmats.2016.48}{ Nat. Rev. Mat. {\bf1}, 16048 (2016)}.

\bibitem{Fong17} P. T. Fong and C. K. Law, Bound state in the continuum by spatially separated ensembles of atoms in a coupled-cavity array, \href{https://doi.org/10.1103/PhysRevA.96.023842}{Phys. Rev. A {\bf96}, 023842 (2017)}.

\bibitem{Calajo19} G. Calaj\'{o}, Y.-L. L. Fang, H. U. Baranger, and F. Ciccarello, Exciting a Bound State in the Continuum through Multiphoton Scattering Plus Delayed Quantum Feedback, \href{https://doi.org/10.1103/PhysRevLett.122.073601}{Phys. Rev. Lett. {\bf122}, 073601 (2019)}.
\bibitem{Facchi19} P. Facchi, D. Lonigro, S. Pascazio, F. V. Pepe, D. Pomarico, Bound states in the continuum for an array of quantum emitters, \href{https://arxiv.org/abs/1904.13004}{	arXiv:1904.13004 (2019)}.

\bibitem{Dinc18} F. Din\c{c}, A. M. Bra\'{n}czyk, I. Ercan, Real-space time dynamics in waveguide QED: bound states and single-photon-pulse scattering, \href{https://arxiv.org/abs/1809.05164}{arXiv:1809.05164v2 (2018)}.
% Entanglement generation in waveguide {Carlos13, }



\bibitem{Carlos13} C. Gonzalez-Ballestero, F. J. Garc\'{i}a-Vidal, and E. Moreno, Non-Markovian effects in waveguide-mediated entanglement, \href{https://doi.org/10.1088/1367-2630/15/7/073015}{New J. Phys. {\bf15}, 073015  (2013)}.


% Dicke, superradiance {Dicke, Haroche, Rehler71, Eberly72}
\bibitem{Dicke} R. H. Dicke, Coherence in Spontaneous Radiation Processes, \href{https://doi.org/10.1103/PhysRev.93.99}{ Phys. Rev. 93, 99 (1954)}.

\bibitem{Haroche} M. Gross, and S. Haroche, Superradiance: An essay on the theory of collective spontaneous emission, \href{https://doi.org/10.1016/0370-1573(82)90102-8}{ Phys. Rep. {\bf 93}, 301 (1982)}.

\bibitem{Rehler71} N. E. Rehler and J. H. Eberly,  Superradiance, \href{https://doi.org/10.1103/PhysRevA.3.1735}{Phys. Rev. A 3, 1735 (1971)}.

\bibitem{Eberly72}J. H. Eberly, Superradiance Revisited, \href{https://doi.org/10.1119/1.1986858}{Am. J. Phys. {\bf40}, 1374 (1972)}.


% Collective emission experiments
\bibitem{Skribanowitz} N. Skribanowitz, I. P. Herman,   J. C. MacGillivray, and M. S.  Feld, Observation of Dicke Superradiance in Optically Pumped HF Gas, \href{https://doi.org/10.1103/PhysRevLett.30.309}{Phys. Rev. Lett. {\bf30}, 309 (1973)}.

\bibitem{Gross76} M. Gross, C. Fabre, P. Pillet, S. Haroche, Observation of Near-Infrared Dicke Superradiance on Cascading Transitions in Atomic Sodium,  \href{https://doi.org/10.1103/PhysRevLett.36.1035}{ Phys. Rev. Lett. {\bf36}, 1035 (1976)}.

\bibitem{Pavolini85} D. Pavolini, A. Crubellier, P. Pillet, L. Cabaret, and S. Liberman, Experimental Evidence for Subradiance,  \href{https://doi.org/10.1103/PhysRevLett.54.1917}{Phys. Rev. Lett. {\bf54}, 1917 (1985)}.

\bibitem{Devoe96} R. G. DeVoe and R. G. Brewer, Observation of Superradiant and Subradiant Spontaneous Emission of Two Trapped Ions, \href{https://doi.org/10.1103/PhysRevLett.76.2049}{Phys. Rev. Lett. {\bf76}, 2049 (1996)}.

\bibitem{Scheibner07} M. Scheibner, T. Schmidt, L. Worschech, A. Forchel, G. Bacher, T. Passow, and D. Hommel, Superradiance of quantum dots, \href{https://doi.org/10.1038/nphys494}{Nat. Phys. {\bf 3}, 106 (2007)}.

\bibitem{Mlynek14} J. A. Mlynek, A. A. Abdumalikov, C. Eichler, and A. Wallraff, Observation of Dicke superradiance for two artificial atoms in a cavity with high decay rate,  \href{https://doi.org/10.1038/ncomms6186}{Nat. Commun. {\bf5}, 5186 (2014)}.

\bibitem{Solano2017} P. Solano, P. Barberis-Blostein, F. K. Fatemi, L. A. Orozco, S. L. Rolston, Super-radiance reveals infinite-range dipole interactions through a nanofiber, \href{http:/doi.org/10.1038/s41467-017-01994-3}{Nat. Commun. {\bf8}, 1857 (2017)}.



\bibitem{Kim2018} J.-H. Kim, S. Aghaeimeibodi, C. J. K. Richardson, R. P. Leavitt, and E. Waks, Super-Radiant Emission from Quantum Dots in a Nanophotonic Waveguide,  \href{https://doi.org/10.1021/acs.nanolett.8b01133}{Nano Letters {\bf{18}}, 4734 (2018)}.

\bibitem{Chen2018} L. Chen, P. Wang, Z. Meng, L. Huang, H. Cai, D.-W. Wang, S.-Y. Zhu, and J. Zhang, Experimental Observation of One-Dimensional Superradiance Lattices in Ultracold Atoms,  \href{https://doi.org/10.1103/PhysRevLett.120.193601}{Phys. Rev. Lett. {\bf120}, 193601 (2018)}.

% Peters limit {Milonni74, Arecchi70}
\bibitem{Milonni74} P. W. Milonni and P. L. Knight, Retardation in the resonant interaction of two identical atoms,  \href{https://doi.org/10.1103/PhysRevA.10.1096}{Phys. Rev. A {\bf10}, 1096 (1974)}.

\bibitem{Arecchi70} F. T. Arecchi and E. Courtens, Cooperative Phenomena in Resonant Electromagnetic Propagation, \href{https://doi.org/10.1103/PhysRevA.2.1730}{  Phys. Rev. A {\bf2}, 1730 (1970)}.


% nonMarkovian collective effects in structured reservoir \cite{Thanopulos19, Vats98, John97, AGT2017PRA, AGT2017PRL}
\bibitem{Thanopulos19} I. Thanopulos, V. Karanikolas, N. Iliopoulos, E. Paspalakis, Non-Markovian spontaneous emission dynamics of a quantum emitter near a MoS2 nanodisk, \href{https://arxiv.org/abs/1904.09264}{	arXiv:1904.09264 (2019)}.
\bibitem{Vats98} N. Vats and S. John, Non-Markovian quantum fluctuations and superradiance near a photonic band edge, \href{https://doi.org/10.1103/PhysRevA.58.4168}{Phys. Rev. A {\bf58}, 4168 (1998)}.
\bibitem{John97}S. John and T. Quang, Collective Switching and Inversion without Fluctuation of Two-Level Atoms in Confined Photonic Systems, \href{https://doi.org/10.1103/PhysRevLett.78.1888}{ Phys. Rev. Lett. {\bf78}, 1888 (1997)}.
\bibitem{AGT2017PRA}A. Gonz\'{a}lez-Tudela and J. I. Cirac, Markovian and non-Markovian dynamics of quantum emitters coupled to two-dimensional structured reservoirs, \href{https://doi.org/10.1103/PhysRevA.96.043811}{ Phys. Rev. A {\bf96}, 043811 (2017)}.
\bibitem{AGT2017PRL} A. Gonz\'{a}lez-Tudela and J. I. Cirac,  Quantum Emitters in Two-Dimensional Structured Reservoirs in the Nonperturbative Regime, \href{https://doi.org/10.1103/PhysRevLett.119.143602}{ Phys. Rev. Lett. {\bf119}, 143602 (2017)}.

% nonMarkovian collective effects with strong coupling 
\bibitem{Daniele19} D. De Bernardis, T. Jaako, and P. Rabl, Cavity quantum electrodynamics in the nonperturbative regime, \href{https://doi.org/10.1103/PhysRevA.97.043820}{ Phys. Rev. A {\bf97}, 043820 (2018)}.

\bibitem{FicekTanas} Z. Ficek, and R. Tana\'{s}, Entangled states and collective nonclassical effects in two-atom systems,  \href{https://doi.org/10.1016/S0370-1573(02)00368-X}{Phys. Rep. {\bf372}, 369 (2002)}.

{\bibitem{footnoteop} Our interaction Hamiltonian is expressed in terms of  symmetrically ordered atomic and field observables. With this choice of ordering, spontaneous emission can be physically interpreted as resulting equally from both  radiation reaction and vacuum fluctuations~\cite{Milonni73, Milonni75}.}

\bibitem{Milonni73}  P. W. Milonni, J. R. Ackerhalt, and W. A. Smith, Interpretation of Radiative Corrections in Spontaneous Emission, \href{https://doi.org/10.1103/PhysRevLett.31.958}{Phys. Rev. Lett. {\bf31}, 958 (1973)}.

\bibitem{Milonni75}P. W. Milonni and W. A. Smith, Radiation reaction and vacuum fluctuations in spontaneous emission, \href{https://doi.org/10.1103/PhysRevA.11.814}{ Phys. Rev. A {\bf11}, 814 (1975)}.

{\bibitem{footnoteg} The coupling $g(\omega)$ between the emitters and the waveguide includes the details of the emitter waveguide interactions such as overlap integrals between  guided modes  of the waveguide and  emitter eigenstates \cite{Klimov2004,Balykin04, FLK, Solano2019}.  }
\bibitem{Klimov2004} V. V. Klimov and M. Ducloy. Spontaneous emission rate of an excited atom placed near a nanofiber, \href{https://journals.aps.org/pra/abstract/10.1103/PhysRevA.69.013812}{ Phys. Rev. A {\bf69}, 013812 (2004)}.
\bibitem{Balykin04} V. I. Balykin, K. Hakuta, Fam Le Kien, J. Q. Liang, and M. Morinaga, Atom trapping and guiding with a subwavelength-diameter optical fiber, \href{https://doi.org/10.1103/PhysRevA.70.011401}{ Phys. Rev. A {\bf70}, 011401(R) (2004)}.
\bibitem{FLK} F. L. Kien, J. Q. Liang, K. Hakuta, and V. I. Balykin, Field intensity distributions and polarization orientations in a vacuum-clad subwavelength-diameter optical fiber, \href{https://doi.org/10.1016/j.optcom.2004.08.044}{ Opt. Commun. {\bf242}, 445 (2004)}.
\bibitem{Solano2019} P. Solano, J. A. Grover, Y. Xu, P. Barberis-Blostein, J. N. Munday, L. A. Orozco, W. D. Phillips, and S. L. Rolston. Alignment-dependent decay rate of an atomic dipole near an optical nanofiber, \href{https://journals.aps.org/pra/abstract/10.1103/PhysRevA.99.013822}{ Phys. Rev. A {\bf99}, 013822 (2019)}.

\bibitem{Asl03} F. M. Asl, and A. G. Ulsoy, Analysis of a System of Linear Delay Differential Equations, \href{https://doi.org/doi:10.1115/1.1568121}{ J. Dyn. Sys., Meas., Control {\bf125}, 215 (2003)}.
\bibitem{footnote1}{Note that for $\phi_p=(2p+1)\pi$ the expressions of the super- and sub-radiant coefficients are exchanged. In general the system evolution is set by both the atomic initial state and the the phase of the EM field acquired upon propagation.}

\bibitem{footnote2}{In the limiting case $d\rightarrow0$ and $\beta = 1$, we obtain  $\tilde c_{\mr{sup}}\bkt{s}\rightarrow{\frac{1}{\sqrt{2}\bkt{s+\gamma}}}$, i.e. $c_\mr{sup}\bkt{t}\rightarrow{\frac{1}{\sqrt{2}} e^{ - \gamma t}}$, and $\tilde c_{\mr{sub}}\bkt{s}\rightarrow\pm{\frac{1}{\sqrt{2}{s}}}$, i.e. $c_\mr{sub}\bkt{t}\rightarrow\pm{\frac{1}{\sqrt{2}}}$, corresponding to the limit of usual Dicke super- and sub-radiance when the emitters are close to each other. Similarly,  if the emitters are far apart, in the limit $d\rightarrow{\infty }$ one obtains  $c_{\mr{sup, sub}}(t) = \frac{1}{\sqrt{2}}e^{-\gamma t/2} $, as both the emitters emit independently.}



\bibitem{Cray82} M. Cray, M.-L. Shih, and P. W. Milonni, Stimulated emission, absorption, and interference, \href{ https://doi.org/10.1119/1.12956}{Am. J. Phys. {\bf50}, 1016 (1982)}.


\bibitem{Corless96}  R. M. Corless, G. H. Gonnet, D. E. G. Hare, D. J. Jeffrey, D. E. Knuth, On the Lambert-$W$ function, \href{https://doi.org/10.1007/BF02124750}{Adv. Comput. Math. {\bf5}, 329 (1996)}.

\bibitem{SM}Supplemental Material contains the details on solution to the atomic dynamics in terms of Lambert-$W$ functions, alternative solution in terms of wavepacket oscillations,  derivation of the field intensity expressions, measure of non-Markovianity, the dynamics of emitter-field correlations, and a summary of experimental feasibility in different platforms.

\bibitem{SPIE} K. Sinha, P. Meystre, P. Solano, \href{https://doi.org/10.1117/12.2530927}{Quantum Nanophotonic Materials, Devices, and Systems 11091 (2019).}

\bibitem{PabloReview}P. Solano, J. A. Grover, J. E. Hoffman, S. Ravets, F. K. Fatemi, L. A. Orozco, S. L. Rolston, Optical Nanofibers: A New Platform for Quantum Optics, \href{https://doi.org/10.1016/bs.aamop.2017.02.003}{ Adv. At. Mol. Opt. Phys. {\bf66}, 439 (2017)}. 
\bibitem{footnote4} We note that for the subradiant state with  $\beta = 1$ when a BIC state is formed, there is a non-zero steady state atomic population in addition to the excitation in field modes. We exclude that special case from consideration here. We have also ignored here the possibility that the emission into external modes can exhibit cooperative effects.

\bibitem{Kurizki15}   G. Kurizki, P. Bertet, Y. Kubo, K. M{\o}lmer, D. Petrosyan, P. Rabl, and J. Schmiedmayer,  Quantum technologies with hybrid systems, \href{https://doi.org/10.1073/pnas.1419326112}{Proc. Natl. Acad. Sci. U.S.A. {\bf112}, 3866 (2015)}. 

\bibitem{Johnson2019} A. Johnson, M. Blaha, A. E. Ulanov, A. Rauschenbeutel, P. Schneeweiss, J. Volz, Observation of Multimode Strong Coupling of Cold Atoms to a 30-m Long Optical Resonator, \href{https://arxiv.org/abs/1905.07353}{	arXiv:1905.07353 (2019)}.

\bibitem{Kato2019} S. Kato, N. N\'{e}met, K. Senga, S. Mizukami, X. Huang, S. Parkins, and T. Aoki, Observation of dressed states of distant atoms with delocalized photons in coupled-cavities quantum electrodynamics, \href{https://doi.org/10.1038/s41467-019-08975-8}{Nat. Commum. {\bf{10}}, 1160 (2019)}.


\bibitem{vanloo2013} A. F. van Loo, A. Fedorov, K. Lalumi\'{e}re, B. C. Sanders, A. Blais, and A. Wallraff, Photon-Mediated Interactions Between Distant Artificial Atoms,  \href{https://doi.org/10.1126/science.1244324}{Science {\bf{342}}, 1494 (2013)}.

\bibitem{Mirhosseini2018} M. Mirhosseini, E. Kim, X. Zhang, A. Sipahigil, P. B. Dieterle, A. J. Keller, A. Asenjo-Garcia, D. E. Chang, and O. Painter, Cavity quantum electrodynamics with atom-like mirrors, \href{https://doi.org/10.1038/s41586-019-1196-1}{Nature {\bf 569}, 692 (2019)}.

% \bibitem{Wang19} D. Wang, H. Kelkar, D. Martin-Cano, D. Rattenbacher, A. Shkarin, T. Utikal, S. G\"{o}tzinger, and V. Sandoghdar, Turning a molecule into a coherent two-level quantum system, 
% \href{https://doi.org/10.1038/s41567-019-0436-5}{Nat. Phys. {\bf15}, 483 (2019)}.

% \bibitem{Turschmann2017} P. T\"{u}rschmann, N. Rotenberg, J. Renger, I. Harder, O. Lohse, T. Utikal, S. G\"{o}tzinger, and V. Sandoghdar, Chip-Based All-Optical Control of Single Molecules Coherently Coupled to a Nanoguide,   \href{https://pubs.acs.org/doi/abs/10.1021/acs.nanolett.7b02033}{Nano Lett. {\bf{17}}, 4941 (2017)}.


\bibitem{Schuurmans}M. F. H. Schuurmans and  D. Polder, Superfluorescence and amplified spontaneous emission: A unified theory, \href{https://doi.org/10.1016/0375-9601(79)90477-8}{ Phys. Lett. A {\bf72}, 306 (1979)}.

\bibitem{Inst} B. Ohayon and G. Ron, New approaches in designing a Zeeman Slower, \href{https://doi.org/10.1088/1748-0221/8/02/P02016}{ J. Inst. {\bf8}, 02016  (2013)}.
\bibitem{Ads} E. Floratos,  G.  Georgiou, and G. Linardopoulos, Large-Spin Expansions of GKP Strings,  \href{https://doi.org/10.1007/JHEP03(2014)018}{J. High Energ. Phys. {\bf 03}, 0180  (2014)}.
\bibitem{Sinha18} K. Sinha, B. P. Venkatesh, and P. Meystre, Collective Effects in Casimir-Polder Forces, \href{https://doi.org/10.1103/PhysRevLett.121.183605}{ Phys. Rev. Lett. {\bf121}, 183605 (2018)}.
\bibitem{Fuchs18} S. Fuchs, and S. Y. Buhmann, Purcell-Dicke enhancement of the Casimir-Polder potential, \href{https://doi.org/10.1209/0295-5075/124/34003}{Eur. Phys. Lett. {\bf 124}, 34003 (2018)}.
\bibitem{Chanda16} T. Chanda, and S. Bhattacharya, Delineating incoherent non-Markovian dynamics using quantum coherence, \href{https://doi.org/10.1016/j.aop.2016.01.004}{Annals of Physics {\bf366}, 1 (2016)}.
\bibitem{Bose2000}  S. Bose, and V. Vedral, Mixedness and teleportation, \href{https://doi.org/10.1103/PhysRevA.61.040101}{ Phys. Rev. A {\bf61}, 040101(R) (2000)}. 


\end{thebibliography}

\begin{thebibliography}{10}
\bibitem{Corless96}  R. M. Corless, G. H. Gonnet, D. E. G. Hare, D. J. Jeffrey, D. E. Knuth, On the Lambert-$W$ function, \href{https://doi.org/10.1007/BF02124750}{Adv. Comput. Math. {\bf5}, 329 (1996)}.
\bibitem{Milonni74} P. W. Milonni and P. L. Knight, Retardation in the resonant interaction of two identical atoms,  \href{https://doi.org/10.1103/PhysRevA.10.1096}{Phys. Rev. A {\bf10}, 1096 (1974)}.

\bibitem{Arecchi70} F. T. Arecchi and E. Courtens, Cooperative Phenomena in Resonant Electromagnetic Propagation, \href{https://doi.org/10.1103/PhysRevA.2.1730}{  Phys. Rev. A {\bf2}, 1730 (1970)}.
\bibitem{Dorner02} U. Dorner and P. Zoller, Laser-driven atoms in half-cavities,  \href{https://doi.org/10.1103/PhysRevA.66.023816}{Phys. Rev. A {\bf66}, 023816 (2002)}.
\bibitem{Kato2019} S. Kato, N. N\'{e}met, K. Senga, S. Mizukami, X. Huang, S. Parkins, and T. Aoki, Observation of dressed states of distant atoms with delocalized photons in coupled-cavities quantum electrodynamics, \href{https://doi.org/10.1038/s41467-019-08975-8}{Nat. Commum. {\bf{10}}, 1160 (2019)}.

\bibitem{Kim2018} J.-H. Kim, S. Aghaeimeibodi, C. J. K. Richardson, R. P. Leavitt, and E. Waks, Super-Radiant Emission from Quantum Dots in a Nanophotonic Waveguide,  \href{https://doi.org/10.1021/acs.nanolett.8b01133}{Nano Letters {\bf{18}}, 4734 (2018)}.

\bibitem{vanloo2013} A. F. van Loo, A. Fedorov, K. Lalumi\'{e}re, B. C. Sanders, A. Blais, and A. Wallraff, Photon-Mediated Interactions Between Distant Artificial Atoms,  \href{https://doi.org/10.1126/science.1244324}{Science {\bf{342}}, 1494 (2013)}.

\bibitem{Mirhosseini2018} M. Mirhosseini, E. Kim, X. Zhang, A. Sipahigil, P. B. Dieterle, A. J. Keller, A. Asenjo-Garcia, D. E. Chang, and O. Painter, Cavity quantum electrodynamics with atom-like mirrors, \href{https://doi.org/10.1038/s41586-019-1196-1}{Nature {\bf 569}, 692 (2019)}.

% \bibitem{Turschmann2017} P. T\"{u}rschmann, N. Rotenberg, J. Renger, I. Harder, O. Lohse, T. Utikal, S. G\"{o}tzinger, and V. Sandoghdar, Chip-Based All-Optical Control of Single Molecules Coherently Coupled to a Nanoguide,   \href{https://pubs.acs.org/doi/abs/10.1021/acs.nanolett.7b02033}{Nano Lett. {\bf{17}}, 4941 (2017)}.
\bibitem{Inst} B. Ohayon and G. Ron, New approaches in designing a Zeeman Slower, \href{https://doi.org/10.1088/1748-0221/8/02/P02016}{ J. Inst. {\bf8}, 02016  (2013)}.
\bibitem{Ads} E. Floratos,  G.  Georgiou, and G. Linardopoulos, Large-Spin Expansions of GKP Strings,  \href{https://doi.org/10.1007/JHEP03(2014)018}{J. High Energ. Phys. {\bf 03}, 0180  (2014)}.
\bibitem{Sinha18} K. Sinha, B. P. Venkatesh, and P. Meystre, Collective Effects in Casimir-Polder Forces, \href{https://doi.org/10.1103/PhysRevLett.121.183605}{ Phys. Rev. Lett. {\bf121}, 183605 (2018)}.
\bibitem{Fuchs18} S. Fuchs, and S. Y. Buhmann, Purcell-Dicke enhancement of the Casimir-Polder potential, \href{https://doi.org/10.1209/0295-5075/124/34003}{Eur. Phys. Lett. {\bf 124}, 34003 (2018)}.
\bibitem{Chanda16} T. Chanda, and S. Bhattacharya, Delineating incoherent non-Markovian dynamics using quantum coherence, \href{https://doi.org/10.1016/j.aop.2016.01.004}{Annals of Physics {\bf366}, 1 (2016)}.
\bibitem{Bose2000}  S. Bose, and V. Vedral, Mixedness and teleportation, \href{https://doi.org/10.1103/PhysRevA.61.040101}{ Phys. Rev. A {\bf61}, 040101(R) (2000)}. 
\end{thebibliography}
\end{document}